\documentclass[a4paper,fleqn,usenatbib]{mnras}
\usepackage{longtable}
\usepackage{lingmacros}
\usepackage{caption}
\usepackage{tree-dvips}
\usepackage[export]{adjustbox}

\usepackage{newtxtext,newtxmath}
\usepackage[T1]{fontenc}
\usepackage{ae,aecompl}
\usepackage{graphicx}	
\usepackage{amsmath}	
\usepackage{amssymb}
\usepackage{soul}
\hypersetup{pdfauthor={Mojgan Aghakhanloo},
               pdftitle={Modelling Luminous-Blue-Variable Isolation},
               pdfkeywords={binaries: general -stars: evolution -stars: massive -stars: variables: general},
               bookmarksnumbered=true}

\newcommand{\msun}{M_{\odot}}

\title[Modelling LBV Isolation]{Modelling Luminous-Blue-Variable Isolation}

\author[Aghakhanloo et al.]{
Mojgan Aghakhanloo,$^{1}$\thanks{ma14g@my.fsu.edu}
Jeremiah W. Murphy,$^{1}$\thanks{jwmurphy@fsu.edu}
Nathan Smith,$^{2}$\and
and Ren\'{e}e Hlo\v{z}ek$^{3}$
\\
$^{1}$Physics, Florida State University, 77 Chieftan Way, Tallahassee, FL 32306, USA\\
$^{2}$Steward Observatory, 933 N.Cherry Ave, Tucson,  AZ 85719, USA\\
$^{3}$Dunlap Institute for Astronomy and Astrophysics, University of Toronto, 50 St. George Street, Toronto, Ontario, Canada M5S 3H4}
\pubyear{2017}
\begin{document}
\label{firstpage}
\pagerange{\pageref{firstpage}--\pageref{lastpage}}

\maketitle
\begin{abstract}
Observations show that luminous blue variables (LBVs) are far more
  dispersed than
massive O-type stars, and \citeauthor{ST15} suggested that these large separations are inconsistent with a single-star evolution model of LBVs. Instead, they suggested that the large distances are most consistent with binary evolution scenarios. To test these
  suggestions, we modelled young stellar clusters and their passive
dissolution, and we find that, indeed, the standard single-star
  evolution model is mostly inconsistent with the
  observed LBV environments. 
If LBVs are single stars, then the lifetimes inferred from their
luminosity and mass are far too short to be consistent with their
extreme isolation.
This implies that there is either an inconsistency in the
  luminosity-to-mass mapping or the mass-to-age mapping.  In this paper, we
  explore binary solutions that modify the mass-to-age mapping and
  are consistent with the isolation of LBVs. For the binary scenarios, our crude models suggest that LBVs are rejuvenated stars. They are either the result
of mergers or they are mass gainers and
received a kick when the primary star exploded. In the merger
scenario,
if the primary is about 19 $\msun$, then the binary has
enough time to wander far afield, merge and form a rejuvenated star.
In the mass-gainer and kick scenario, we find that LBV isolation is
consistent with a wide range of kick velocities, anywhere from  0 to $\sim$ 105 km/s. In either scenario, binarity seems to play a major role in the
isolation of LBVs.
\end{abstract}

\section{INTRODUCTION} \label{INTRODUCTION}
Stellar mass is one of the primary characteristics that determine a
star's evolution and fate \citep{W15}; therefore, understanding
mass-loss is important in developing a complete theory of stellar
evolution. Yet, understanding the physics and relative importance of
steady and eruptive mass-loss in the most massive stars remains a
major challenge in stellar evolution theory.  There has been
substantial progress in understanding mass-loss via steady line-driven
winds of hot stars \citep{K00, P08}, and this effect is included in
stellar evolution models \citep{VD01, W02, MM05, MP13}.  However, the
mass-loss rates of red supergiants (RSGs) and the role of eruptive mass-loss
remain unclear, and the influence on stellar evolution remains
uncertain \citep{SO06,S14}. The luminous blue variable (LBV)
is one such poorly constrained class of eruptive stars.

LBVs are luminous, unstable massive stars that suffer irregular
variability and major mass-loss eruptions \citep{HD94}. The mechanism
of these eruptions and the demographics of which stars experience
these is poorly constrained \citep{S11,S14}.  The traditional view
has been that most stars above 25-30 $\msun$ pass through an LBV
phase in transition from core H burning to He burning. In this brief phase, they experience eruptive mass-loss as a means to transition from a
hydrogen-rich star to an H-poor Wolf--Rayet (WR) star \citep{HD94}. In
this scenario, LBVs experience high mass-loss due to an unknown
instability, which may be driven by a high luminosity-to-mass (L/M) ratio, near the
Eddington limit \citep{HD94, V12}. However, a high L/M ratio may not be sufficient to explain LBV eruptions. Instead, the instability may require rare circumstances such as binary interactions \citep{ST15}.

\citet{ST15} noted that LBVs are isolated, and they proposed that binary interaction is important in LBV evolution and gives rise to their isolation. If LBVs mark
a brief transitional phase at the end of the main sequence and before
core-He burning WR stars, then they should be found near other massive
O-type stars.  However, \citet{ST15} found that LBVs are quite
isolated from O-type stars, and even farther away from O stars than
the WR stars are. Given their isolation, \citet{ST15} concluded that
the LBV phenomenon is inconsistent with a single-star scenario and is
most consistent with binary scenarios. In some respects,
  there was already earlier evidence that the simple LBV-to-WR-to-SN
  mapping is not entirely accurate \citep{S07, S08}. For example, \citet{KV06} proposed an LBV and supernova (SN) connection. \citet{KV06} suggested that modulations in the radio light curve of SNe 2003bg and 1998bw reflected variations in the mass-loss rate similar to S Dor variations. In other cases, some Type IIn SNe may have LBV-like progenitors based on pre-SN mass-loss
  properties (mass, speed, H composition). For example, \citet{O13} reported a pre-supernova outburst 40 d before the Type IIn supernova SN 2010mc. Even though the progenitor of SN 2010mc was not directly identified as an LBV such an outburst is consistent with rare giant eruptions of LBVs. However, there has never been a direct connection between LBVs and Type IIn SNe. Instead, the connection is circumstantial in that narrow lines of Type IIn imply significant mass-loss from the progenitor, and even when the progenitor has been observed to vary, there are generally not enough observations to definitively classify a progenitor as an LBV. On the other hand, the isolation of directly identified LBVs provides a stronger constraint on their evolution \citep{ST15}.
    
In this paper, we constrain whether single-star or binary models are required to explain LBV isolation. We do this by developing
simple models for the dispersal of massive
stars on the sky. Our model is general and we designed it to have very few parameters. This simplicity and generalizability enable us to constrain the spatial and dynamic distributions of many stellar types. In this paper, we focus this generalized approach to model spatial distributions of early, mid and late O-type stars and most importantly LBVs. In particular, we use our models to constrain whether LBV isolation is consistent with single-star evolution or binary evolution.

Part of the reason that LBVs are poorly understood is that there
are few examples. There are only 10 unobscured in our Galaxy and 19
known in the nearest galaxies, the Large Magellanic Cloud (LMC) and Small Magellanic Cloud \citep[SMC;][]{ST15}. Even this
small sample includes `candidate' LBVs (see below).  Classifying
various stars as LBVs or candidates can be somewhat controversial
\citep{HD94, W03,V12}; here we summarize their basic
characteristics. LBVs are luminous, blue massive stars with irregular
or eruptive photometric variability.  Stars that resemble LBVs in
their physical properties and spectra, but lack the tell-tale
variability, are usually called `LBV candidates'.  The reason they
are sometimes grouped together is that it is suspected that the LBV
instability may be intermittent, so that candidates are temporarily
dormant LBVs \citep{S11,S14}. The LBV candidates in \citet{ST15} have shell nebulae that are thought to be indicative of past eruptive mass-loss.

Although the signature eruptive variability of LBVs was identified
long ago, the physical theory of LBV eruptions is not yet clear. For
the most part, LBVs seem to experience two classes of eruptions: S
Doradus (or S Dor) eruptions (1--2 mag) and giant eruptions ($\ge$ 2
mag).

S Doradus variables take their namesake from the prototypical LBV S
Doradus \citep{V01}. During S Dor outbursts, LBVs make transitions
in the HR diagram (HRD) from their normal, hot quiescent state to
lower temperatures (going from blue to red). In its quiescent state,
an LBV has the spectrum of a B-type supergiant or a late Of-type/WN
star \citep{W77,BW89}. In this state, LBVs are fainter (at visual
wavelengths) and blue with temperatures in the range of 12000--30000 K \citep{HD94}.  In their maximum visible state, their spectrum
resembles an F-type supergiant with a relatively constant temperature
of $\sim8000$ K.  S~Dor events were originally proposed to occur at
constant bolometric luminosity \citep{HD94}. So a change in
temperature implies a change in the photospheric radius, $L=4\uppi
\sigma R^2 T^4$. \citet{HD94} suggested that the eruption is so
optically thick that a pseudo-photosphere forms in the wind or
eruption. However, quantitative estimates of mass-loss rates
show that they are too low to form a large enough pseudo-photosphere
\citep{D96, G09}.  Similar studies also imply that the bolometric
luminosity is not strictly constant \citep{G09}.  Instead, it has been
suggested that the observed radius change of the photosphere can be a
pulsation or envelope inflation driven by the Fe opacity bump
\citep{G12}.

The other distinguishing type of variability is in the form of giant
eruptions like the 19th century eruption of $\eta$ Car \citep{S11}.
The basic difference from S Dor events is that giant eruptions show a
strong increase in the bolometric luminosity and are major eruptive
mass loss events, whereas S Dor eruptions occur at roughly constant
luminosity and are not major mass-loss events.  The mass-loss rate at
S Dor maximum is of the order of $10^{-4} M _{\odot}yr^{-1}$ or less \citep{W89,G09}.
On the other hand, giant eruption mass loss rate is of the order of
$10^{-1}$--$1 \ M _{\odot}yr^{-1}$ \citep{O04,SO06,S14}.  It is
unlikely that a normal line-driven stellar wind is responsible for the
giant eruptions because the material is highly dense and optically
thick \citep{O04,SO06}. Instead, giant eruptions must be
continuum-driven super-Eddington winds or hydrodynamic explosions
\citep{SO06}.  Both of these lack an explanation of the underlying
trigger; the super-Eddington wind relies upon an unexplained increase
in the star's bolometric luminosity, whereas the explosive nature of
giant eruptions would require significant energy deposition.  There is
much additional discussion about the nature of LBV giant eruptions in
the literature \citep{HD94,O04,SO06,S11,S14}.
  
\citet{ST15} highlighted a result that changes the emphasis on the most
likely models.  They found that compared to O stars, LBVs are isolated
in the Milky Way and the Magellanic Clouds. Moreover, they found that
LBVs appear to have a much larger separation than even WR stars, which
are thought to be the descendants of LBVs. They concluded
  that the single-star model is inconsistent with the statistical
  properties of LBV isolation.  At a minimum, they suggested that LBV
  isolation may require binary evolution for a large fraction of LBVs
  if not all.

\citet{H16} put forth a different interpretation of LBV locations,
suggesting that they do not rule out the single-star scenario. They
noted that the sample in \citet{ST15} is a mixture of less luminous
LBVs, more luminous classical LBVs and unconfirmed LBVs, and they
proposed that separating them alleviates the conflict with single-star
models. From their point of view, the single-star hypothesis still
works because (1) the three most luminous stars of the sample that are
classical LBVs (with initial masses greater than 50 $\msun$) have
a distribution similar to late O-type stars, and (2) the less luminous
LBVs (with initial mass $\sim$25-40 $\msun$) are not associated with
any O stars, but have a distribution similar to RSGs, which could be consistent with them being
single stars on a post-RSG phase. They strongly suggested that one separates the LBVs into two categories by luminosity for future statistical tests. Moreover, \citet{H16} criticized that five of the LMC stars (R81,
R126, R84, Sk-69271 and R99) are neither LBVs nor candidates.

However, \citet{S16} showed that even using the LBV sample subdivided
as \citet{H16} preferred does not change the result that LBVs are too
isolated for single-star evolution (overlooking the lack of
statistical significance).  The most massive LBVs appear to be
associated on the sky with late O-type dwarfs (point 1 above), which,
however, have initial masses less than half of the presumed initial
masses of the classical LBVs.  Similarly, the lower luminosity LBVs
have a similar distribution to RSGs, but these RSGs are dominated by
stars of 10--15 $\msun$ \citep{S16}.

\citet{H16} also stated that the observed LBV  velocities seem to be
too small to be consistent with the kicked mass-gainer scenario, but
\citet{S16} pointed out that without a quantitative model for the
velocity distributions, it would be difficult to rule anything in or
out. In this paper, we will show that both high- and low-luminosity LBVs
and LBV candidates have larger separations than one would expect, and
in Section~\ref{kick velocity} and Fig.~\ref{Kick} we show that a wide
range of kick velocities are consistent with the large separations.

The main goal of this paper is to quantitatively constrain whether the
relative isolation of LBVs is inconsistent with a single-star
evolution model. We begin by reproducing and verifying \citet{ST15}
results (Section~\ref{OBSERVATIONS}). In Section~\ref{SPATIAL_DISTRIBUTION}, we
introduce a simple model for young stellar clusters and their passive
dissolution. To test this model, we also compare the separations
between O stars for the model and observations; we find that the model
reproduces some general properties of the spatial distribution of
massive stars, but it is lacking in other ways.  Since we constructed
the simplest model possible, this implies that we may improve the
dispersion model and learn even more about the evolution of massive
stars. Then in Section~\ref{Cluster dissolution}, we present the primary
consideration of this paper; we compare single-star evolution and
binary evolution in the context of cluster dissolution, and we find
that the single-star evolution scenario is inconsistent for initial
masses appropriate for LBV luminosities.  We discuss two binary
evolution channels that are consistent with the relative isolation of
LBVs. We then summarize, and we discuss future observations to further
constrain the binary models (Section~\ref{SUMMARY}).

\section{OBSERVATIONS}
\label{OBSERVATIONS} 

In the following sections, we explore which theoretical models are most consistent with the data,
but before that, we clearly define, analyse and characterize the
data in this section. First, we
reproduce and verify the results of \citet{ST15}. Secondly, we further characterize the
data, noting that the distributions of nearest neighbours are
lognormal.  Since lognormal distributions have very few
parameters, this restricts the complexity and parameters of our
models in Section~\ref{SPATIAL_DISTRIBUTION}.

\citet{ST15} found that LBVs are much more isolated than O-type or
  WR stars, suggesting that LBVs are not an intermediary stage
  between these two evolutionary stages. In particular, they found that on average, the distance from
 LBVs to the nearest O star is quite large (0.05 deg). For comparison,
 the average distance from early O stars to
 the nearest O star is 0.002 deg, and from mid and late O stars are 0.008 and 0.010 deg
 respectively. If single early- and mid-type O stars are
   indeed the main-sequence progenitors of LBVs, then one would expect
 the spatial separations between LBVs and other O stars to be not too
 different from the separation between early- and mid-type O stars.
 However, the LBV separations are an order of magnitude farther than
 the early- and mid-type separations. In fact, the LBV separations are five times
 larger than even the late-type O stars, which live longer and can
 in principle migrate farther. 
 
\citet{ST15} quantified the difference in the distributions of
separations by using the Kolmogorov--Smirnov (KS) test. Comparing the distributions of LBVs to early, mid and late types gives $P$-values of 5.5e--9, 1.4e--4 and
4.4e--6, respectively. These values imply that O-stars and LBV distributions are quite different. If true, then these results have profound
  consequences for our understanding of LBVs and their place in
  massive star evolution. In fact, \citet{ST15} suggested that the most natural
  explanation is that LBVs are the result of extreme binary
  encounters. Later we will test this assertion, but for now
    we reproduce and verify
their results.

To verify the results of \citet{ST15}, we first define the data. The data 
consist of two main parts: LBVs and O stars. Their sample
includes WR stars, sgB[e] stars and RSGs too, but we do not discuss
them here because at the moment, we want to keep our models in Sections~\ref{SPATIAL_DISTRIBUTION} and \ref{Cluster dissolution} simple and we will focus just on LBVs and O stars. Their LBV samples include 16 stars in the LMC, and
three stars in the SMC. They did not consider Milky Way LBVs
because the distances and intervening line-of-sight extinction in the plane of the Milky Way are uncertain \citep{SS17}. In their study, they
included LBV candidates with a massive CSM (circumstellar medium) shell that
likely indicates a previous LBV-like giant eruption. LBVs and their
important parameters are summarized in Table \ref{tab:lbv}. 

The masses of LBVs that are in Table \ref{tab:lbv} are
  uncertain; specifically, the uncertainty in masses due to distance uncertainties is at least 8\%, but the systematic uncertainties due to the stellar evolution modelling are likely much larger. Currently, it is difficult to adequately quantify these uncertainties. None of them are kinematic mass measurements. Rather,
  they are based upon inferring the mass by comparing their colour and magnitude in the HRD with evolutionary tracks of various masses. In this modelling, the two main sources of uncertainties
  are modelling the uncertain physics of late-stage evolution and
  distance.  The distance uncertainty to the LMC is 3--4\% \citep{MC05, W12, K14}, which would
  translate to a luminosity uncertainty of 6--8\%.  However, the systematic
uncertainties in modelling LBVs and their luminosities and colour are
unknown and could easily be much larger than the distance
uncertainty.  Therefore, like \citet{ST15}, we merely report rough
estimates for the LBV masses in Table \ref{tab:lbv}.

\begin{table}
\caption{List of LBVs and LBV candidates adapted from
   \citet{ST15}. For the stars in the SMC, we rescale their angular
    separation by 1.2 as if they are located at the distance of the
    LMC. Parentheses in the name represent LBV candidates and
    parentheses in mass column specify the LBVs with relatively poorly
    constrained luminosity and mass.}
\label{tab:lbv}
\begin{tabular}{lccr} 
\hline
LBV (name) & Galaxy (name) & $S$ (deg) & $M_{eff}$ ($\msun$)\\
\hline
R143 & LMC  &0.00519 & 60  \\
R127& LMC  &0.00475&90 \\
S Dor& LMC  &0.0138 &55 \\
R81& LMC  &0.1236 &(40)  \\
R110& LMC  &0.2805 &30  \\
R71& LMC  &  0.4448&29 \\
MWC112& LMC  & 0.0892&(60) \\
R85& LMC  &0.0252&28 \\
(R84)& LMC  &0.1575&30 \\
(R99)& LMC  & 0.0412&30 \\
(R126)& LMC  & 0.0358&(40)  \\
(S61)& LMC  &0.1432  &90 \\
(S119)& LMC  &0.3467&50    \\
(Sk-69142a)& LMC  &0.0522&60  \\
(Sk-69279) & LMC  &0.0685&52 \\
(Sk-69271)& LMC  &0.040&50\\
HD5980 & SMC & 0.0191&150 \\
R40 & SMC & 0.1112&32 \\
(R4) & SMC & 0.0160&(30)\\
\hline
\end{tabular}
\end{table}
\citet{ST15} gathered the positions of O-type stars within $10^{\circ}$ projected radius of 30 Dor from SIMBAD data base. They also used the revised Galactic O-star
Catalog \citep{M13} to check their O-star
samples (not shown in their paper), but as they claimed this did not
change their overall results. We collect the same O-star samples from SIMBAD data base. 

\begin{figure}
\includegraphics[width=\columnwidth]{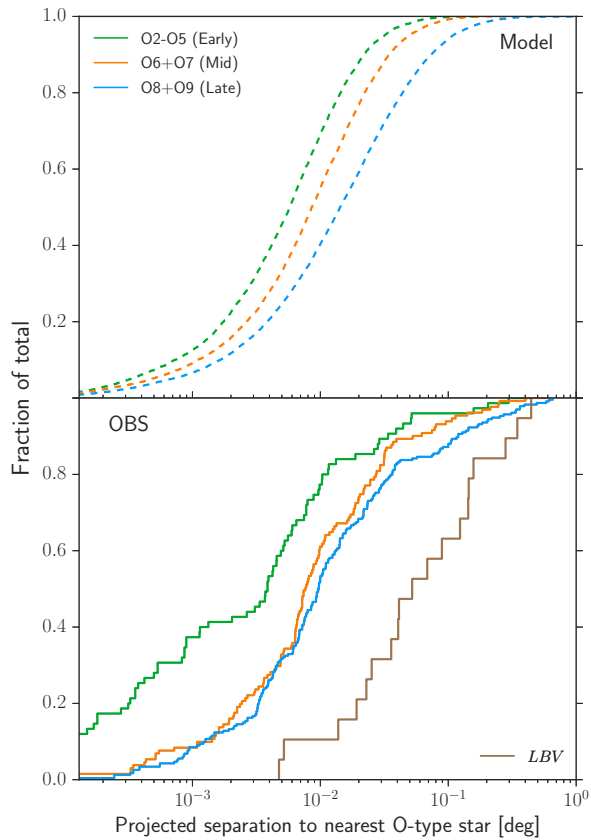}
\caption{Cumulative distributions for the projected
    separation to the nearest O star. The top panel represents the
  modelled distribution for O stars and the bottom panel represents the data for both O stars and LBVs. Later, we will use the modelled O-star distributions to devise a general dispersion model, which we use to model the LBV separations (see Section~\ref{Cluster dissolution}). Broadly, the model reproduces the observations; both show a lognormal distribution, and the average separation
  increases with spectral-type because the later spectral type last longer.}
\label{CumulativeDistribution}
\end{figure}

After gathering the data, we find the distance from one star to the nearest O star. The bottom panel of Fig.~\ref{CumulativeDistribution} shows the
  resulting cumulative distributions (the top panel shows results of our modelling, which we discuss in Section~\ref{SPATIAL_DISTRIBUTION}); note that the
  distributions for LBV and O-type separations are quite distinct. For example, the $P$-value for the comparison of the LBV and the mid-type distributions is $2.8 \times 10^{-5}$. Our
KS-test $P$-values are listed in Table \ref{tab:pvalues}. We consider three KS tests. In one, we compare the separation for both confirmed and candidate LBVs with O-star distribution. In the second, the LBV distribution only includes confirmed LBVs, and in the third, the LBV distribution only contains the candidates. When we include both
confirmed and candidate LBVs, the LBV and O-star distributions are clearly not drawn from the same parent
  distribution. However, omitting LBV candidates reduces the
  distinctions between the distributions.  One might argue that since LBVs represent a
    later evolutionary stage, then the spatial separations should be
    larger, and therefore, the distributions of early-type O stars and
  LBVs should not represent the same distribution.
However, we will show in section~\ref{Cluster dissolution} that the lifetimes of massive stars
    are far too short to explain these large discrepancies. In our initial assessment, we agree with
    \citet{ST15}; the large separations present a challenge to the
  single-star evolution scenario. In the next sections, we will
  present theoretical models to quantify this inconsistency.

\begin{table}

\caption{The $P$-values for KS tests for the distributions of separation. We are comparing the separations between LBVs and O stars and the separations between O stars of various types. Broadly, we reproduce the results of \citet{ST15} 
who found that the distribution of separations between LBVs and the
nearest O star is quite different from the distributions for the
separations between O stars and the nearest O star.
The second row shows the results of our KS tests
between the LBV separations and the early-, mid-
and late-type O stars.  Our results are similar to
those of \citep[first row]{ST15}, first row. Like \citet{ST15} 
we obtain the positions of O stars and their rayet
spectral types from SIMBAD. \citet{ST15} 
updated the spectral types with the 
Galactic O-star Catalog \citep{M13}; 
however, we did not. This slight difference 
in spectral typing is what causes the modest 
difference in $P$-values. In either case, the 
LBV separations are inconsistent with 
any O-type separations. If we exclude 
the LBV candidates (third row), the conclusions remain the same, but the
significance is greatly reduced.}
\label{tab:pvalues}
\resizebox{\linewidth}{!}{
\begin{tabular}{lccr} 
\hline
Data set & Early O & Mid O& Late O\\
\hline
LBV+LBVc \citep{ST15}& 5.5e--9&1.4e--4&6.4e--06 \\
LBV+LBVc (this work)&8.2e--08& 2.8e--05 &8.4e--05\\
LBV (this work)&9.2e--04 &2.3e--02&5.7e--02\\
LBVc (this work)& 6.4e--06  &5.1e--05&2e--04\\
\hline
\end{tabular}
}
\end{table}

\begin{figure}
\includegraphics[width=\columnwidth]{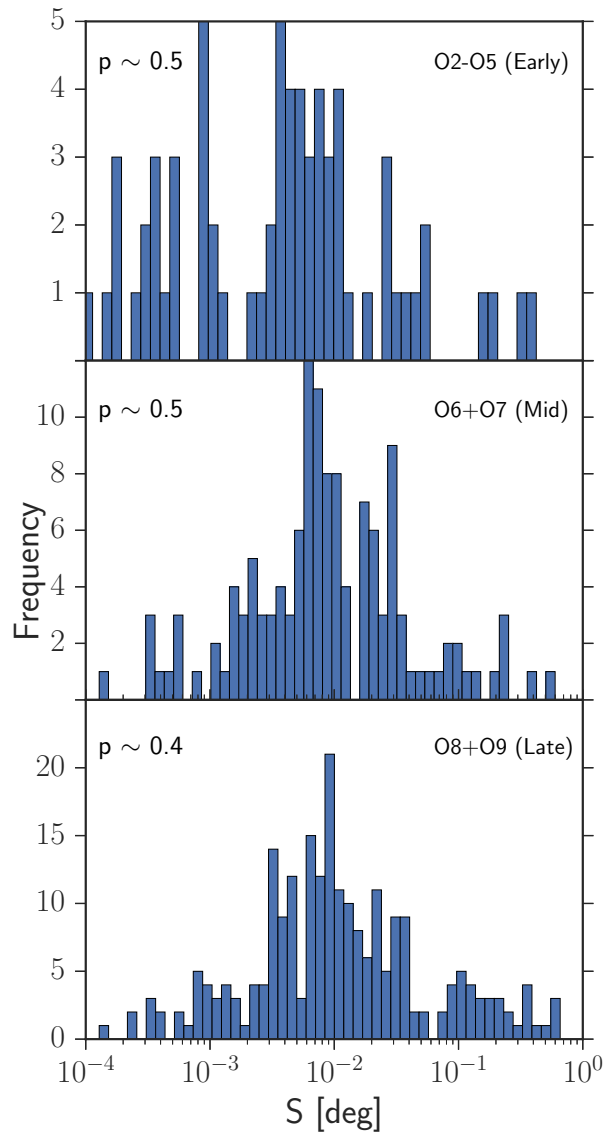}
\caption{Normality test. The distributions of separations for early, mid,and late O stars are consistent with a lognormal distribution. In each plot, we show the probability, $p$, that the parent distribution is a lognormal distribution.}
\label{Normality}
\end{figure}

Before we constrain the models, note that the separation distributions
  are lognormal (see Fig.~\ref{Normality}).  In fact, this simple
  observation greatly restricts the complexity of the models that we
  may explore in the next sections.  If a variable such as the
  separation between stars shows a lognormal distribution, then there are
  only two free parameters that describe the distribution, the mean
  and the variance.  In addition, if the separation depends upon other
variables such as a velocity distribution, then thanks to the
central limit theorem, the separation distribution will only depend
upon the mean of the variance of the secondary variables such as the
velocity distribution.  This means that we cannot propose overly
complex models for the velocity distribution.  We would only be able
to infer the mean and the variance anyway.  Fortunately, we may
measure the separation for different types of O stars and other
evolutionary stages.  This means that we may infer the temporal evolution in
addition to the mean and variance.
Whatever models we
propose, they cannot be too elaborate; we will only be able to infer the mean and variance of one
quantity as a function of time.

\section{A GENERIC MODEL FOR THE SPATIAL DISTRIBUTION OF THE STARS IN A PASSIVE DISPERSAL CLUSTER}\label{SPATIAL_DISTRIBUTION}
In order to model the relative
  isolation of LBVs, we need to model the dissolution of clusters and
  associations of massive stars. 
For several reasons, we model the dissolution of young stellar clusters with a
minimum set of parameters. For one, the O-star distributions are
lognormal. Therefore, there
  are only a few parameters that describe the data that one may fit.
The only data that we can reliably fit
  are the mean, variance and time evolution of the separations. So
whatever models we develop, they should not be overly 
complex. Also, as
far as we know, there are no simple self-consistent and tested models for the
dissolution of clusters. Therefore, we propose a simple model of
  cluster dissolution and adapt it to consider two scenarios: cluster
  dissolution in the context of single-star evolution and cluster
  dissolution with close binary interactions. In this section, we present a cluster dispersal model considering only single-star evolution. While our dissolution models represent the
    spatial distributions reasonably well in certain respects, we note that our model fails to match the data in other ways.  This implies that our model is
  missing something.  In other words, we may be able to infer more
  physics about the dissolution of clusters from the simple spatial
  distribution of O stars. In the next section, we contrast the single-star model with a model that considers binarity.

Our main goal is to introduce a model for young stellar clusters that
predicts the spatial distribution of massive stars,
  especially O stars. We start by considering the simplest model. In
  the following, we model the average distance to the nearest O star by nothing
  more than the passive dispersal of
  a cluster. 
  
  Before we dive into the details
    of the model, it is worth characterizing the scales of a typical
    cluster. We begin right after star formation
    ends and consider a system of gas and stars that is in virial equilibrium. In this case, we have 
$2T+U=0$, where $T$ is the total thermal plus kinetic energy and $U$ is the
gravitational potential energy. Initially, the system with total
  mass $M_{T}$ and radius $R$ is bound, and the stars have a velocity dispersion that scales as the gravitational potential of the entire system $\sigma_v \sim (GM_{T}/R)^\frac{1}{2}$.
 Then the system loses gas mass by some form
of stellar feedback (UV radiation, stellar winds, etc.) and likely makes the stars unbound. If
  the system loses all of the gas quickly, then the stars will drift
  away with a speed roughly equal to the velocity dispersion when the
  cluster was bound.  Hence, $v_d \sim \sigma_v$. All that is left to do is estimate $M_{T}$ and $R$. A typical cluster has $R\sim 4$ pc and about 40 O stars; if only $\sim 1\%$
of the gas in giant molecular clouds form stars
\citep{K07}, the total mass of the molecular cloud, $M_{T}$, is the order of  $2 \times 10^5 \msun$. Given these approximations, we estimate that the drift velocity is the order of $v_{d}\sim 13.5 (\frac{M_{T}}{2 \times 10^5 \msun}\frac{4 pc}{R})^\frac{1}{2}$.

Next, we present a more specific dissolution model to convert
    this dispersal velocity into a distribution of separations as a
    function of time.  Rather than using this estimate for the
    dispersal velocity, we will use the data and our model to infer
    the dispersal velocities. We
  propose a Monte Carlo model for the dissolution of the
  clusters. First, we randomly sample $N_{cl}$ clusters
  uniformly in time between 0 and 11 Myr. For each
    cluster, we draw a cluster mass from a distribution of cluster
    masses. Then,  we estimate the total number
  of the stars ($N_{*}$), and for each cluster, we
    draw a distribution of stellar masses ($M_{*}$) from the Salpeter
    distribution.

First, we randomly select a total number of O stars
, $N_{*}$, for each cluster. The
  distribution from which we draw the size of each cluster is the
  Schechter function \citep{E97}, 
  $\frac{dN_{cl}}{dM_{cl}} \propto M_{cl}^{-2}$, where $M_{cl}$ is the
  mass of the cluster.  However, we are most interested in the number
  of O stars for each cluster, so our first order of business is to
  express the Schechter function in terms of the number of O stars.
The mass of the cluster is $M_{cl}= A
\int_{M_{*1}}^{M_{*2}} {M_{*}}^{-1.35}$, where $M_{*1}$
  and $M_{*2}$ are the minimum and maximum masses of O star that we
  consider. In terms of this, the total number of O stars becomes
\begin{equation}
\label{eq:52}
N_{*}= \frac{M_{cl}}{\int_{M_{*1}}^{M_{*2}} {M_{*}}^{-1.35} {dM_{*}}}  \int_{M_{*1}}^{M_{*2}} {M_{*}}^{-2.35} {dM_{*}} \, .
\end{equation}
Therefore, the total number of 
stars in the cluster is proportional to the mass of the cluster
($N_{*}\propto {M_{cl}}$), and we can easily translate the
  distribution in mass to a distribution in the number of stars for
  each cluster,
  $\frac{dN_{cl}}{dN_{*}} \propto N_{*}^{-2}$. If $R_{*}$
  is drawn from the uniform distribution between 0 and 1, then the total number of stars in the cluster is
\begin{equation}
\label{eq:60}
N_{*}= \frac{1}{R_{*}({N_{*max}}^{-1}-{N_{*min}}^{-1})+{N_{*min}}^{-1}} \, ,
\end{equation}
where $N_{*max}$ and $N_{*min}$ are the maximum and minimum number of
the stars in the cluster.

For each star, we draw the mass from the Salpeter initial
  mass function (IMF),
\begin{equation}
\label{eq:63}
M_{*}=(\frac{1}{[R_m({M_{max}}^{-1.35}-{M_{min}}^{-1.35})]+{M_{min}}^{-1.35}})^{0.74} \, ,
\end{equation}
where $R_{m}$ is a random number between 0 and 1.

Having established the initial conditions, we now describe the evolution.
The average separation between O stars
  depends upon how much the cluster has dispersed and how many O stars
are left.  So we need to model the dispersion of the O stars and their
disappearance. Therefore, we need to model the spatial distribution (or spatial density) and time evolution of massive stars in a cluster. Once we establish the spatial distribution, we then calculate the separation between stars. The distribution of separations in essence is a convolution of the density function with itself. Because this is a multiplicative process, the central limit theorem implies a lognormal distribution. The central limit theorem also dictates that any underlying spatial distribution with a well-defined mean and variance results in a lognormal distribution. Therefore, we are free to choose a simple model for the spatial distribution, and we choose a Gaussian for the spatial distribution.

For the time evolution we assume that each cluster is passively dispersing with a typical velocity scale of $v_{d}$.Therefore, the characteristic size scale of the Gaussian spatial distribution is $\sigma=v_{d}t$. Given the assumption
that stars are coasting then the individual velocities
are $r/t$. With
  these assumptions, then the distribution of velocities is Gaussian too, $p(v)=\frac{t}{\sqrt{2\uppi\sigma^2}}e^{-r^2/2\sigma^2}$.
  
\begin{figure}
\includegraphics[width=\columnwidth]{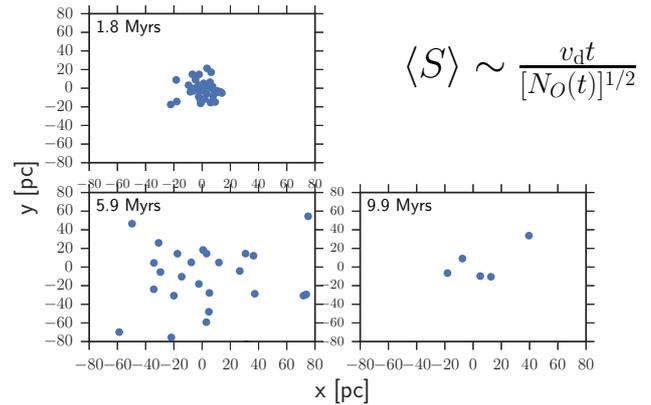}
\caption{We propose a Monte Carlo model for the separations between O stars and LBVs by considering a random sample of dissolving clusters at random ages. Here we show the O stars of three
 randomly generated clusters, each with its own age. Note that the average separation between the O stars increases with age for two reasons. First, the separations increase as the cluster disperses with a drift velocity $v_d$ over time $t$. Secondly, O stars disappear as they evolve.}
\label{SpatialDistribution}
\end{figure}

Another important aspect of modelling these clusters is to model the age and disappearance of massive stars. For the lifetimes, we use the results of single-star
 evolutionary models from the binary population synthesis code, \textsc{binary\_c} \citep{I04,I06,I09}. Therefore, the average separation between stars goes up both because the cluster is dispersing and O stars are disappearing. Fig.~\ref{SpatialDistribution}
shows the spatial distribution of an example model at several ages.

With our model defined,
our first task is to constrain whether the average distances between LBVs and O
stars are consistent with the passive dissolution of a cluster with
single-star evolution. To compare our models to the data, we
   calculate the angular separations, assuming that the clusters are
   at the distance of the LMC. Furthermore, to be consistent with \citet{ST15}, we
   subdivide the modelled O stars into early, mid and late types based
upon their masses. To convert from mass to spectral type,
we used \citet{M05} data. Early-type O stars have masses greater than
34.17 $\msun$, late-type O stars have masses $\le 24.15$ $\msun$ and mid-type
  O stars have masses in between. In the next subsection, we
test whether our passive single-star dissolution model is consistent
with the data.

\subsection{COMPARING THE PASSIVE SINGLE-STAR DISSOLUTION MODEL WITH THE DATA}

Next, we compare the passive dissolution of single stars to the
  LMC and SMC nearest-neighbour distributions. Fig.~\ref{CumulativeDistribution} shows the cumulative distribution for the
  separations for our simple dissolution model (top panel) and for the
observations (bottom panel). For illustration purposes, we set $v_d$ to
14.5 km/s, making the modelled distribution have about the
same mean as the data. So far, our passive
dissolution model is in good agreement with observations. Both the model and observations show a
  lognormal distribution in separations, and the average separation
  increases with spectral type, which is expected since
  later O stars live longer and have more time to disperse.

\begin{figure}
\includegraphics[width=\columnwidth]{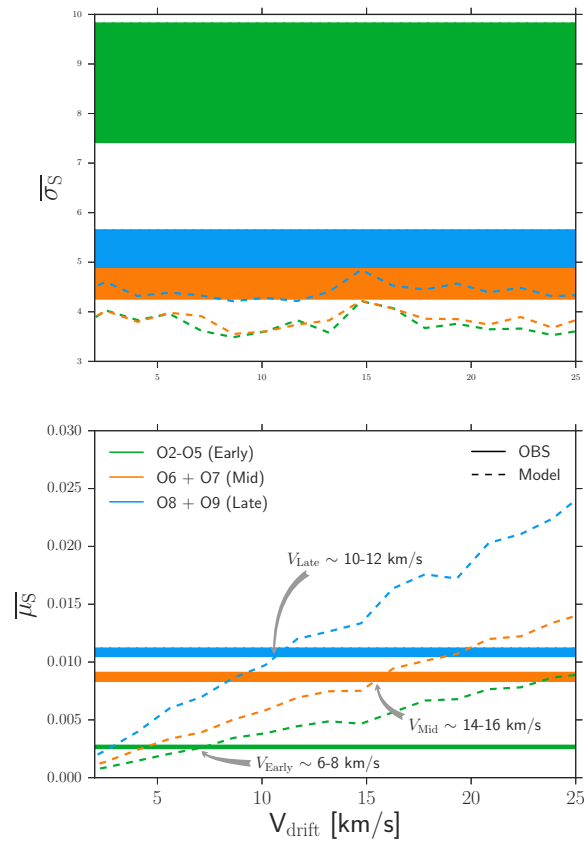}
\caption{The
  mean (bottom panel) and std. deviation (top panel) distance to the
  nearest neighbour versus drift velocity. We calculate the mean and std. deviation in log first; then, we calculate the  $\overline{\mu_{S}}=10^{\mu \ (log \ S)}$ and $\overline{\sigma_{S}}=10^{\sigma \ (log \ S)}$}. In both panels, dashed lines
  represent the passive dissolution model and solid lines
  represent the observational data \citep{ST15}. We highlight three
  main conclusions. (1) The drift
    velocities that we infer by comparing our simple model with the
    data are roughly what we estimated in Section~\ref{SPATIAL_DISTRIBUTION}. (2) We infer
    larger drift velocities for the later type O stars, implying that
    binary evolution and kicks may be important. (3) The passive
    dissolution model is not able to reproduce the variance in the
    distributions, which implies missing physics from our model. In other words, there is room to improve our model and learn more about the interplay between O-star evolution and cluster dissolution.
\label{MeanVariance}
\end{figure}

Because the distributions are lognormal, there are only two parameters that describe the distribution, the mean and std. deviation. Therefore, we investigate how our model reproduces these two distribution characteristics. The primary parameter in our model
  is $v_d$, so in Fig.~\ref{MeanVariance} we plot the mean (bottom panel) and
  std. deviation (top panel)
as a function of $v_d$. The dashed
  lines represent the modelled mean and std. deviation, and the solid
  bands indicate the observed values. The vertical axes in Fig.~\ref{MeanVariance} are $\overline{\mu_{S}}$ and $\overline{\sigma_{S}}$. First, we calculate the mean and std. deviation in log; then, we calculate $\overline{\mu_{S}}=10^{\mu (log \ S)}$ and $\overline{\sigma_{S}}=10^{\sigma (log \ S)}$.
The solid bands provide some estimate of uncertainty in our inferred drift velocity, we bootstrap
the observations, giving a variance for both the mean and
std. deviation.

We draw three main
conclusions from Fig.~\ref{MeanVariance}.  For one, the drift
velocities that we infer by comparing our simple model with the
data are roughly what we would expect; see our order-of-magnitude
estimate in Section~\ref{SPATIAL_DISTRIBUTION}.  Secondly, we
  infer larger drift velocities for
  the late-type O stars (10--12 km/s) in comparison to early-type stars
  (6--8 km/s). However, this trend is not monotonic; the mid-type O
  stars have an inferred drift velocity (14--16 km/s) that is similar to but slightly
  higher than the late-type O stars. Thirdly, our simple model is not able to
  reproduce the variance in the distributions.  This implies that
  something is missing from our model. In other words, there is more
  that we can learn about the evolution of massive stars in clusters
  from their spatial distributions. Despite the shortcomings, the model is able to reproduce the average separations with reasonable drift velocities. Therefore, we proceed with our analyses under these caveats.

Since the early-type O stars are
  more massive and have lower velocities, it is natural to consider
  mass segregation as the reason for these lower velocities.  However,
  the relaxation time is of the order of 100 Myr, which is more than
  the maximum age of late-type O stars (11 Myr). So, it is unlikely
  that these systems have enough time to reach equipartition and mass
  segregation. Despite this fact, we test this idea and
  we find that the inferred velocities are not readily consistent with
  equipartition anyway. In equilibrium, the stars in a cluster are in
  equipartition in their kinetic energies.  Therefore, the ratio of
  masses for two stars should equal the inverse ratio squared of their
  velocities: $m_{i}/m_{j}=(v_{j}/v_{i})^2$.
  Comparing late to early, the ratio of masses is
  $m_{late}/m_{early}\sim 0.3$ and the ratio of the squared
  velocities is $(v_{early}/v_{late})^2\sim 0.4$.  This seems
  consistent with mass segregation.  However, the other comparisons do
  not.  For mid and early, $m_{mid}/m_{early}\sim 0.43$ and
  $(v_{early}/v_{mid})^2\sim 0.21$, which is a factor of 2
  off.  The late-to-mid comparison gives $m_{late}/m_{mid}\sim
  0.7$ and $(v_{mid}/v_{late})^2\sim 1.85$, which is also a
  factor of 2 off. Furthermore, if equipartition in kinetic energy
  were valid, then all of these ratios should have similar values.
  We have yet to adequately assess the uncertainties in these ratios;
  that will take significant more modelling.  Even so, the fairly large
  discrepancies seem to rule out kinetic energy equipartition in
  the cluster.

We can use the results in Fig.~\ref{MeanVariance} to also infer that LBV
  isolation puts interesting constraints on their evolution.  The
  average separation for late-type O stars is 0.01 deg.  For LBVs,
  the average separation is roughly five times bigger.  Dimensionally, the
average separation should be proportional to the dispersion velocity
and the age, $S \sim v_d t_{\rm age}$. If an LBV
  comes from the most massive stars, then one would not expect them to
have ages larger than the late-type O stars. Therefore, as a conservative
estimate, let us assume that an LBV is an evolved massive star
that has about the same age as a late O-type
star.  Under this assumption, since the separations for LBVs are five
times bigger than late-type O stars, this implies that the dispersal
velocity is five times bigger than the late-type O star, which is of the order of 100 km/s. To be more quantitative, in the next sections, we extend the
passive model to infer the actual dispersal velocity for LBVs.
Alternatively, we consider binary scenarios that may give an explanation for
the relatively large isolation for LBVs.


\section{Cluster dissolution with close binary interactions}\label{Cluster dissolution}

In the previous section, we suggested that the
single-star dispersal model is inconsistent with the
  isolation of LBVs. In this section, we put the passive
    single-star dispersal model to the test, and show that it is
    indeed inconsistent with observations.  In addition, we consider models that involve binary interactions in
a dispersing cluster. Our aim is to
develop models to see whether binary scenarios are consistent with the
LBV observed separations. At the moment, there is
  very little information other than the separations, so it is not
  worth developing an overly complex model for binary interaction.  We
would not be able to constrain the extra parameters of the model.
Therefore, we develop the simplest binary models to constrain the
data. In particular, we consider two simple models
that involve binary evolution in a dispersing cluster. In the first
model, we consider that an LBV is the product of a merger and
is a rejuvenated star; in the second model, we
consider that an LBV is a mass gainer, which would also
  be a rejuvenated star, and receives a kick when
its primary companion explodes. See Fig.~\ref{MergerCartoon} and~\ref{KickCartoon}.

In Section~\ref{PASSIVE MODEL}, we first put together
  an analytic model for the average separation between
  two stars versus time. Then in Section~\ref{inconsistency passive} we use
  this model to show the inconsistency in the single-star
  model, and we show that LBVs are either overluminous given their
  mass or they are the product of a merger and are a rejuvenated
  star.  Alternatively, in Section~\ref{KICK MODEL}, we use the analytic
  model to develop a kick model, and in Section~\ref{kick velocity}, we use
  this model to infer a potential kick velocity for LBVs. In
    summary, we illustrate
that the isolation of LBVs is consistent with binary
 scenario and is inconsistent with the single-star model.
 
\begin{figure}
\includegraphics[width=\columnwidth]{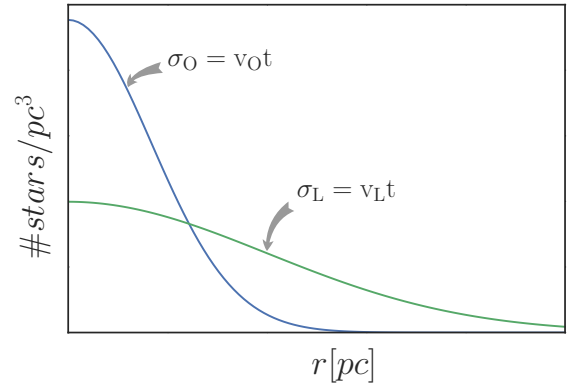}
\caption{Two simple spatial-distribution models for the derivation of our analytic scalings.}
\label{Display}
\end{figure}

To constrain the models, we first derive analytic
  scalings for the average separations, and then we explore whether
  these scalings are consistent with simple binary models.
First, we consider
simple models for the spatial distribution of two groups of stars,
type O and type L.  Each has a Gaussian
spatial distribution with its own velocity dispersion
$v_d$ ,which we label as $v_{\rm O}$ and $v_{\rm L}$.
Later, we will consider two
  scenarios: one in which these average velocity dispersions are the
same, and one in which they are different. For a visual representation of these simple
  models, see Fig.~\ref{Display}.  Given these
distributions, we calculate the average separation between a
star and the nearest star in the same group. Then we calculate the average separation between a star
in group O and a star in group L. The average separation between
stars in the same population is 
 \begin{equation}
  \label{eq:11}
 \langle S \rangle= 2\uppi \int_{0}^{\infty} S \ p(r) \ rdr \, ,
 \end{equation}
 where $S$
is the separation, and $p (r)$ is the probability density function
$p(r)=\frac{1}{2\uppi\sigma^2}e^{-r^2/2\sigma^2}$ where $\sigma=vt$. To calculate the
mean value of the separation, we need to find the separation ($S$). One
way to estimate the distance to the nearest
neighbour is to use the spatial density of stars. In general, an estimate for the
  distance to the nearest neighbour is, $\hat{S} \approx 1/n^{1/d}$,
  where $n$ is the number density and $d$ is the number of dimensions that we consider \citep{Z14}. When viewing clusters projected on to
  the sky, $d = 2$,
   \begin{equation}
  \label{eq:12}
\hat{S} \approx 1/n^{1/2} \, .
 \end{equation}
 If we consider a simple density distribution, $n(r) = \frac{N}{2 \uppi
  \sigma^2} \exp{-r^2/2\sigma^2}$, then the average separation in two-dimensional space is
 \begin{equation}
  \label{eq:10}
\langle \hat{S} \rangle=2 \left ( \frac{2\uppi}{N} \right )^{1/2}\sigma \, ,
\end{equation}
where $N$ is the total number of stars in the cluster.

Now we consider two populations of stars. One we
represent with `O' , which represents
the largest number of tracer
stars. As the label suggests, we will later consider O stars
  as a large number of tracer stars. The other,
  `L', represents a more rare set of tracer stars, which may have a
  different density distribution than the first. Obviously, later `L'
  will represent LBVs.  In this
case, the combined density is 
 \begin{equation}
 \label{eq:16}
n_{OL}(r)=\frac{N_{O}}{2\uppi\sigma_{O}^2}e^{-r^2/{2\sigma_{O}^2}}+\frac{N_{L}}{2\uppi\sigma_{L}^2}e^{-r^2/{2\sigma_{L}^2}} \, .
\end{equation}

With this two-component
  expression for the density, we can evaluate the local separation,
$\hat{S}$, via equation.~(\ref{eq:12}) and then we can
calculate the average separation from
equation.~(\ref{eq:11}). Calculating the average separation is
  numerically straightforward.  However, with a small but useful
  assumption, we can derive an analytic estimate for the average
  separation. To make it easier to calculate the integral analytically,
  we make two assumptions. First, we assume that the average separation is roughly given by the scale of one over the square root of the average density. Therefore, 
\begin{equation}
 \label{eq:13}
\langle S_{OL}
\rangle \approx \frac{1}{{\langle n_{OL}
    \rangle}^{1/2}} \, .
    \end{equation} 
Secondly, because LBVs are extraordinarily rare compared to the O stars, we assume that $N_{O} \gg N_{L}$. By considering these two assumptions, the average density is
\begin{equation}
 \label{eq:23}
\langle n_{OL} \rangle \approx \frac{N_{O}}{2\uppi(\sigma_{O}^2+\sigma_{L}^2)} \, .
\end{equation}
Once we plug this into the equation for ${\langle n_{OL}
    \rangle}$, equation.~(\ref{eq:13}) leads to the average distance
from LBVs to the nearest O star:
\begin{equation}
 \label{eq:143}
\langle S_{OL}
\rangle \approx \left (\frac{2\uppi(\sigma_{O}^2+\sigma_{L}^2)}{N_{O}(t)}
\right )^{1/2}
    \end{equation} 
    
Soon we will use the separation between O
stars to help constrain the models for LBVs, so we now derive an
analytic model for $\langle S_O \rangle$.  The average density for O
stars is 
\begin{equation}
\langle n_O \rangle = \frac{N_O(t)}{4 \uppi \sigma_O^2} \, ,
\end{equation}
and so the average separation between O stars is roughly
\begin{equation}
\langle S_O \rangle \approx \frac{1}{\langle n_O \rangle^{1/2}}
= \left ( \frac{4 \uppi }{N_O(t)} \right )^{1/2} \sigma_O \, .
\end{equation}

To make use of these expressions for the average separation, we need
to compare the separations between two different tracer populations.
Because the masses of mid-type O stars correspond roughly to inferred
minimum mass of LBVs, we use the mid-type O-star average separation as
a reference:
\begin{equation}
\label{eq:analytic1}
\left ( \frac{\langle S_{OL} \rangle}{\langle S_O \rangle} \right )^2 =
\frac{1}{2} \left ( 1 + \frac{\sigma_L^2}{\sigma_O^2}\right )
\frac{N_O(t_O)}{N_O(t_L)} \, ,
\end{equation}
where we are careful to consider how the
number of O stars changes with time and we evaluate this function at
the age of the LBV population and the reference O-star population.
This equation represents the general expression relating age, the
average separations and the drift (or kick) velocity of each tracer
population.

In the expressions for the average separations, the separations grow
due to two effects: a drift velocity and the death of O stars.  The
drift part is simply proportional to $t$.  Next, we explicitly derive
the number of O stars as a function of time, $N_O(t)$.Given a
  mass function $dN/dM$, the total number of O stars is
$N_{O}=\int_{M_{1}}^{M_{2}}\frac{dN}{dM}dM=\frac{A}{-\alpha+1}(M_{2}^{-\alpha+1}-M_{1}^{-\alpha+1})$,
where $M_1$ is the minimum mass for an O star ($\sim$16 $\msun$), $M_2$ is
the maximum mass for an O star, which is a function of the age of the cluster,
$\alpha$ is the slope (we use Salpeter, 2.35) and $A$ is a
normalization constant. If
  we assume a power-law relationship between mass of an O star and its
lifetime as an O star, then we can relate the age of an $M_2$ O star
to the age of an $M_1$ O star:
$M_{2}=M_{1}(\frac{t}{t_{1}})^{-1/\beta}$. From the binary
  population synthesis code, \textsc{binary\_c} \citep{I04,I06,I09}, we find that
the value of $\beta \sim 1.7$. Combining these
expressions, we get an equation for the number of O stars as a
function of the age of the cluster, $t$,
\begin{equation}
 \label{eq:34}
N_{O}(t)=\frac{A}{1-\alpha}\left(1-(\frac{t}{t_{1}})^\tau\right){M_{1}}^{1-\alpha}
\, .
\end{equation}

With an explicit function for the number of O stars, we may now derive
the equation relating separation, age and velocity, including
explicitly all of the dependence on time.  Substituting the expression
for $N_O(t)$, equation.~(\ref{eq:34}) into the general analytic expression,
equation.~(\ref{eq:analytic1}), we finally arrive at the general analytic
formula, explicitly relating separation, age and velocity:
\begin{equation}
\label{eq:analytic2}
\left ( \frac{\langle S_{OL} \rangle}{\langle S_O \rangle} \right )^2 =
\frac{1}{2} \left ( 1 + \left ( \frac{v_L x}{v_O x_O} \right )^2 \right )
\frac{(1 - x_O^\tau)}{(1 - x^\tau)} \, ,
\end{equation}
where $\tau={\frac{\alpha-1}{\beta}}$, $x_{O}=\frac{t_{O}}{t_{1}}$ and
$x=\frac{t_{L}}{t_{1}}$. $t_{1}$ (11 Myr) is
 the age of the minimum mass and $t_O$ (3 Myr) is a reference
  age. We estimate these values from binary population
    synthesis code, \textsc{binary\_c} \citep{I04,I06,I09}. 

In the next subsections, we
use our general analytic result, equation.~(\ref{eq:analytic2}), to explore what average separations one would
expect when we consider the passive dissolution in three scenarios, a
single-star evolution scenario, a binary scenario that involves a
merger and a binary scenario that involves a kick.

\subsection{PASSIVE MODEL}
\label{PASSIVE MODEL}
Using our analytic estimates for the average separation, we
  assume that the dispersal velocities for LBVs and O stars are the same
and estimate the average separation for the passive single-star
model. Comparing this model to the observations,
we find that the
  passive single-star model is inconsistent
with the observations. If LBVs do passively disperse
  with the same velocity as the rest of the O stars, then we propose
  that LBVs are the product of a merger and are rejuvenated stars.
  
In this case, we need to consider the average separation when the
dispersal velocities for LBVs and O stars are the same. In this
scenario, our general analytic expression, equation.~(\ref{eq:analytic2}), reduces to 
\begin{equation}
 \label{eq:38}
\frac{S_{L}}{S_{O}}=\left [ \frac{1}{2}
\left ( 1 + \left (\frac{x}{x_O}\right )^2 \right )
\frac{(1-{x_{O}}^\tau)}
     {(1-x^\tau)} \right ]^{1/2} \, .
\end{equation}
This equation represents the passive model.

\subsection{INCONSISTENCY IN THE PASSIVE MODEL IMPLIES MERGER AND REJUVENATION}
\label{inconsistency passive}
\begin{figure*}
\includegraphics[width=1.75\columnwidth]{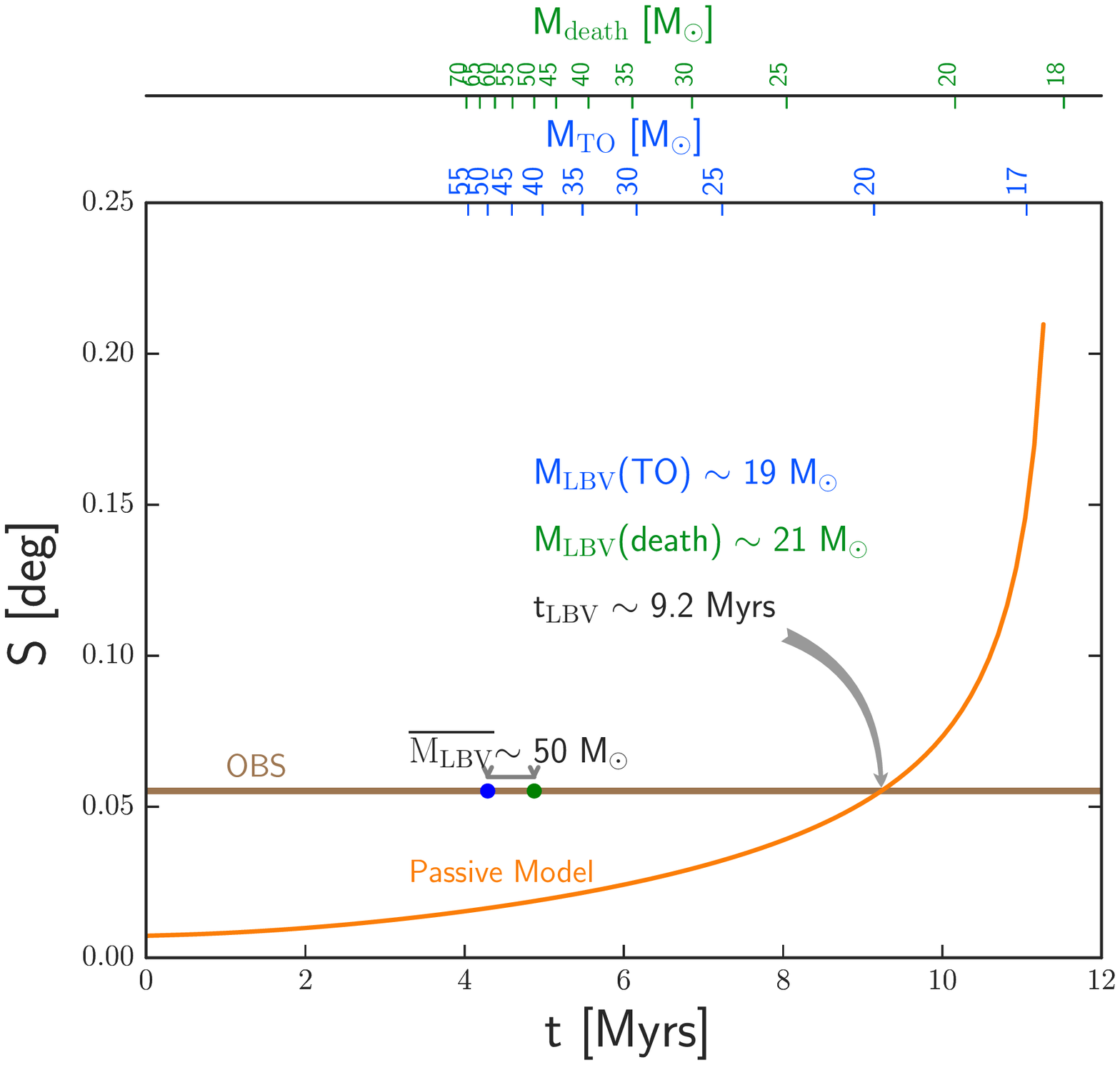}
  \caption{The relative
    isolation of LBVs is consistent with a binary merger in which the LBV is a rejuvenated star.
  The solid brown line shows the observed LBV average separation. If LBVs passively dissolve with the
    rest of the cluster (orange curve),
    then we infer an average age for LBVs of 9.2 Myr.  This
    corresponds to the main-sequence turn-off time for a 19 $\msun$ star
    and the death time for a 21 $\msun$ star.  However, the average mass
  for LBVs estimated from their luminosities is roughly 50 $\msun$.
  Stars this massive do not live long enough to passively disperse to
  large distances.  On the other hand, if LBVs are the products of a
  merger, and the primary has a mass between about 19 and 21 $\msun$, then the rejuvenated star could have a high luminosity, high mass and
  old age allowing it to disperse to larger distances.}
  \label{Merger}
\end{figure*}

\begin{figure}
\includegraphics[width=\columnwidth]{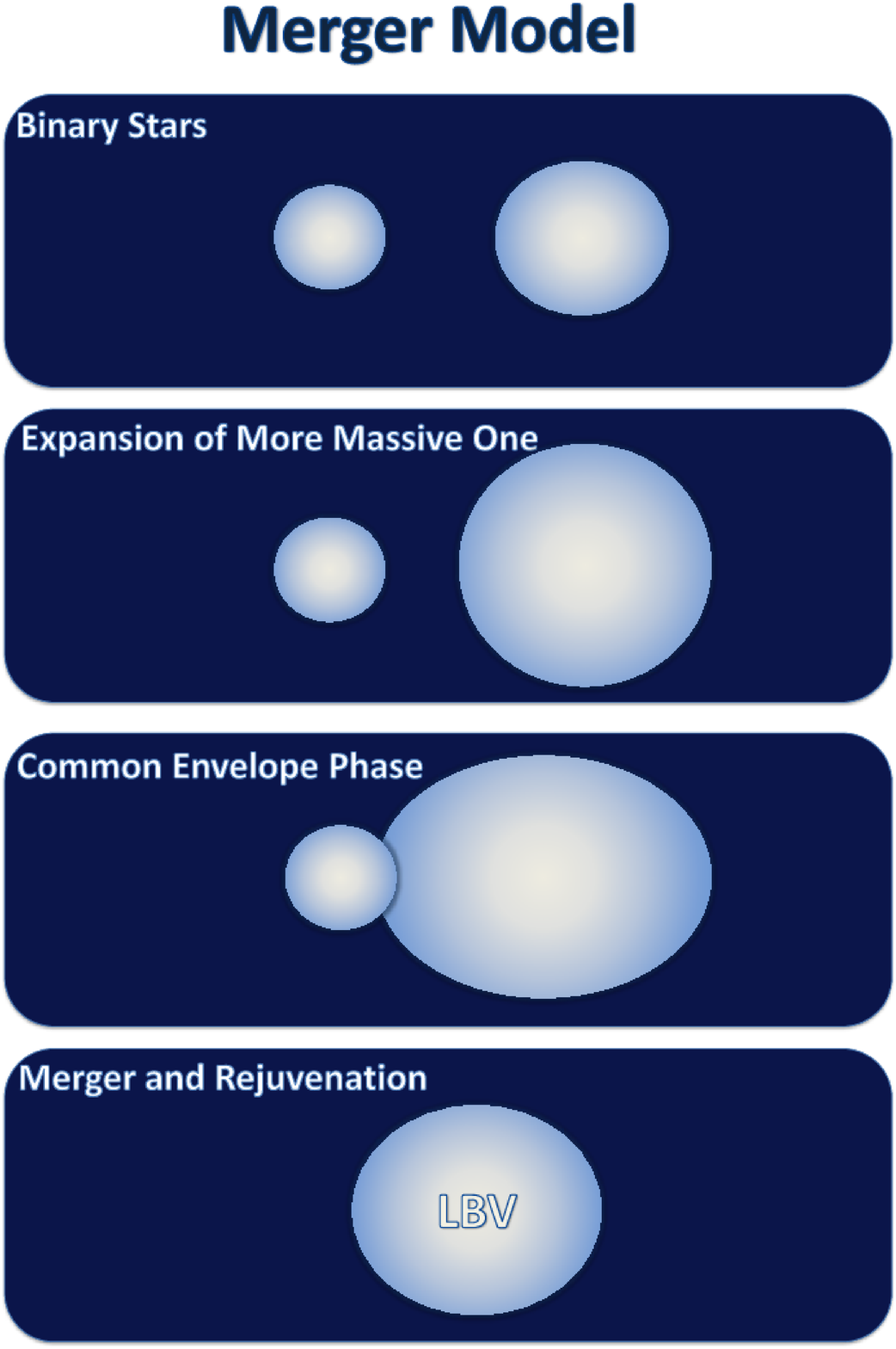}
\caption{Merger model outline. In this binary scenario, LBVs are a
  product of rejuvenation of two massive stars. For a given mass, a
  rejuvenated star has a larger maximum possible age than a
  single-star counterpart. These larger maximum ages allow a
  rejuvenated star enough time to drift farther from other O
  stars. This is one binary scenario that is consistent with the isolation of LBVs.}
\label{MergerCartoon}
\end{figure}

Next, we use the passively
  dissolving solution, equation.~(\ref{eq:38}), to show that the isolation of LBVs is
  inconsistent with the single-star scenario.  If LBVs are massive
  stars above 21 $\msun$ and evolve as isolated stars, then Fig.~\ref{Merger} and Fig.~\ref{MergerCartoon}
  demonstrate that the maximum ages of these LBVs are wholly
  inconsistent with the large separations observed for massive stars.

\begin{figure*}
\includegraphics[width=1.75\columnwidth]{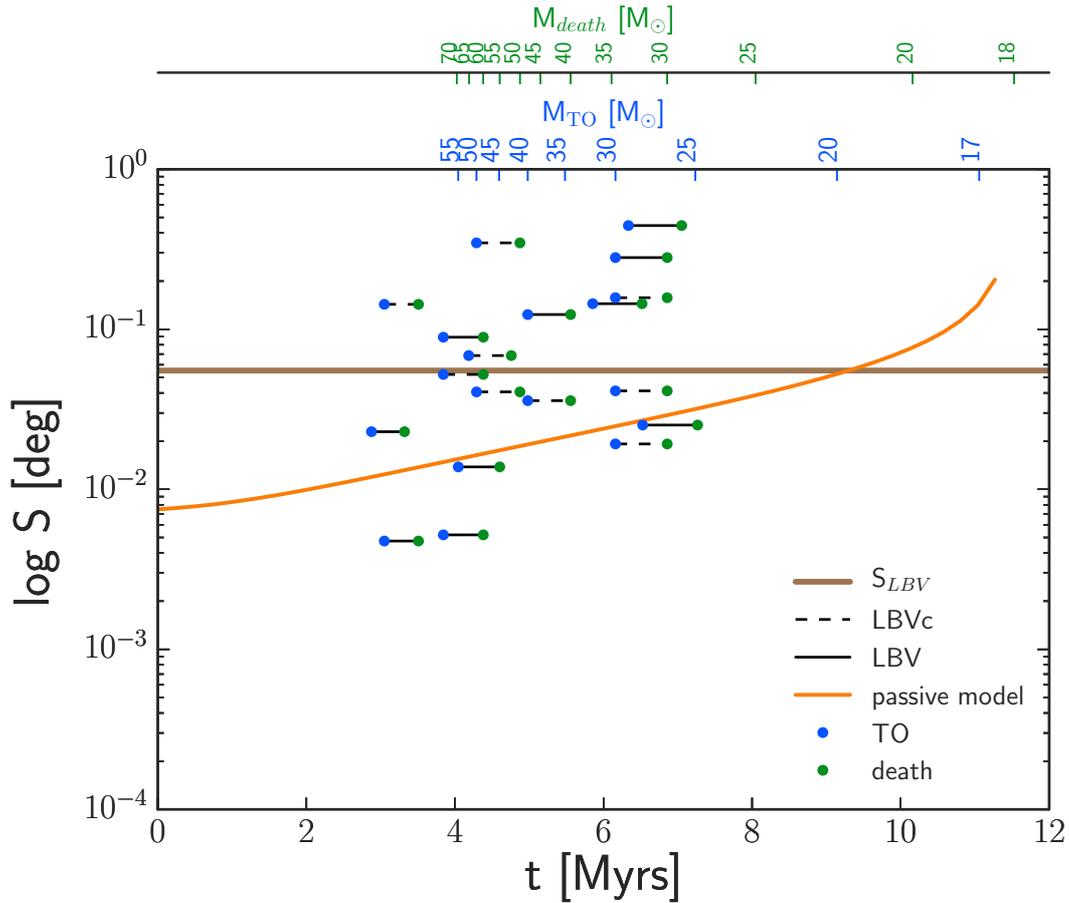}
  \caption{LBV isolation
     is inconsistent with the single-star model and passive
     dissolution of the cluster. The solid brown line shows the LBVs'
     average separation obtained from the data. The orange curve shows
     our analytic description, equation.~(\ref{eq:38}), for the average
     separation in the context of passive dissolution.  This model
     requires a reference; we used the mid-type O observations as the
     reference.  The line segments show the purported masses and
     allowable ages for the LBVs (solid segments) and LBV candidates
     (dashed segments).  If LBV mass estimates are correct, then even
     when one considers the maximum age for LBVs, the separations are
     much larger than what the single-star passive model predicts.}
  \label{Passive}
\end{figure*}

The passive model predicts much lower separations than the observational data. See Fig.~\ref{Passive} for an illustration. The orange curve represents the passive model, $S_{LBV}$ in equation.~(\ref{eq:38}). The solid brown line illustrates the LBVs' average separation obtained from the data compared to the reference average separation, $\frac{S_{LBV}}{S_{0}}$. It is clear that most of the LBVs have larger separations compared to what the passive model predicted.

Moreover, in the passive model, LBVs do not have enough time to get to
the observed average separation. Fig.~\ref{Merger} shows
  the same passive model, the observed LBV separation, but this time
  we simplify the possible ages of LBVs by showing the ages for the
  average mass of our LBV sample.  Clearly, if LBVs evolve as a normal
single star, then they do not have enough time to reach the large
separations.  Instead, let us consider how old an LBV would have to be
in order to passively disperse to the observed separations.
Fig.~\ref{Merger} shows that the age would need to be about 9.2 Myr.
Yet this age corresponds to the main-sequence turnoff time for a 19 $\msun$ star or the death of a 21
  $\msun$ star. Both of these values are below the
average mass of the LBVs, 50 $\msun$
(Section~\ref{OBSERVATIONS}). It is clear that considering LBVs in the
context of a standard single-star evolution is inconsistent with the
isolation.  

In short, the luminosity-to-age
mapping of single-star models is inconsistent with the extreme
isolation of LBVs.  One can consider this mapping in two steps: an
age-to-mass mapping and a mass-to-luminosity mapping.Technically, the breakdown in the
  luminosity-to-age mapping could be a result of the breakdown in
  either one of these steps.  In other words, LBVs could be far more
  luminous than their masses would suggest.  At the moment, there is
  no known physics that would lead to this, so we instead consider how
  binary evolution may alter the mass-to-age mapping.

Assuming that the drift velocities
of the LBVs and the O stars are the same, then one possible solution is that LBVs are the result of 
mergers and are rejuvenated stars. See Fig.~\ref{MergerCartoon} to
visualize the merger model. We are not the first to
  suggest that LBVs are linked to close binary interaction. For
  example, see \citet{J14} and \citet{G89}. What is different here is that, following \citet{ST15}, we analyse how the spatial distribution of LBVs strongly suggests close binary interactions.

\subsection{KICK MODEL}
\label{KICK MODEL}
\begin{figure}
\includegraphics[width=\columnwidth]{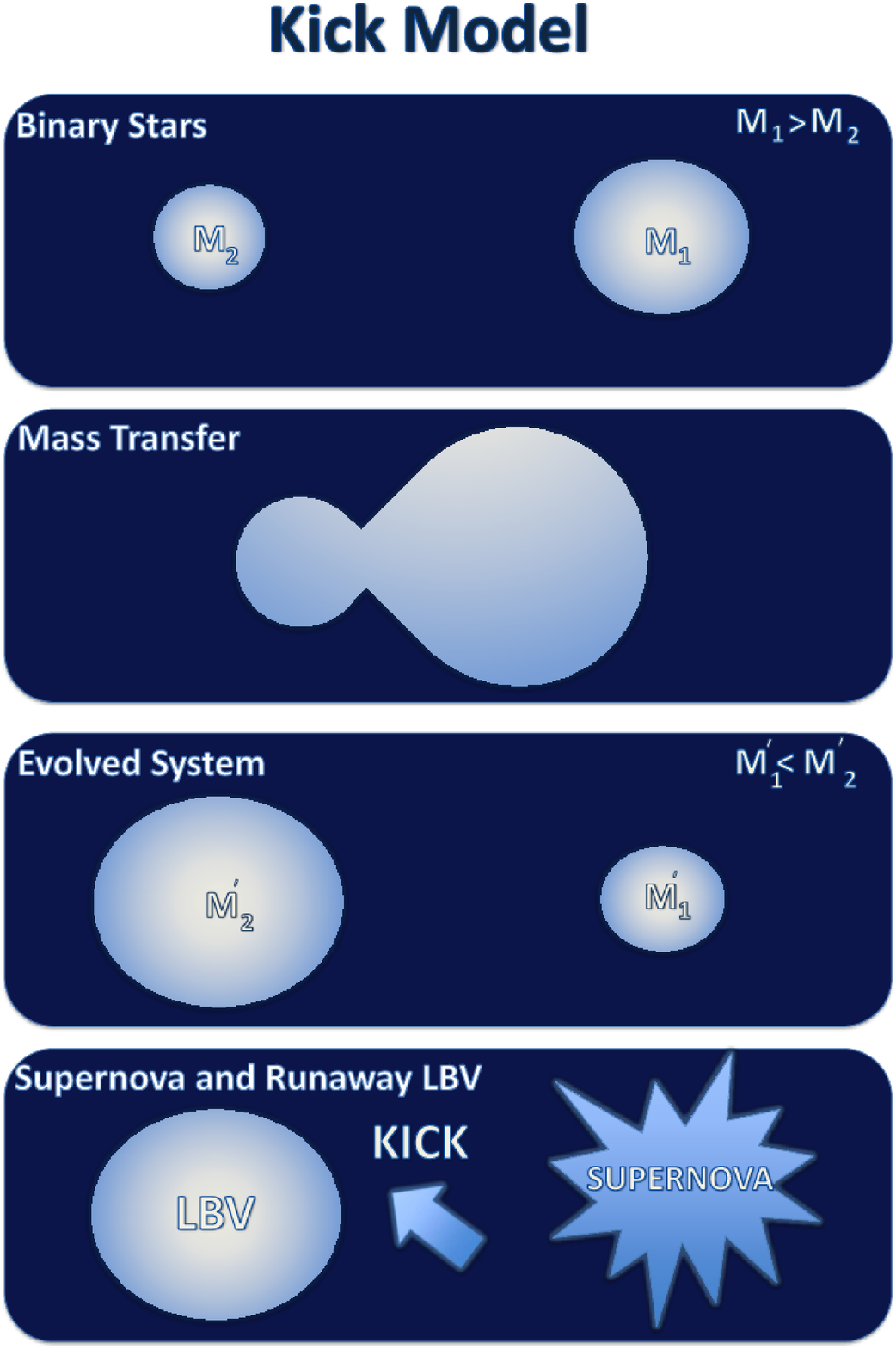}
\caption{Kick model outline. In this binary scenario, a pre-LBV star
  gains mass from its more massive companion star. After mass
  transfer, the mass gainer (LBV) receives a kick when its companion
  explodes in an SN.}
\label{KickCartoon}
\end{figure}

Another binary model that is consistent with the isolation of 
LBVs is the kick model.
To visualize
  the kick model, consider the binary scenario in Fig.~\ref{KickCartoon}. In this
model, the primary star, the more massive star, evolves first
  and transfers mass to the secondary star. If the more
  massive star is massive enough to explode as a core-collapse
  SN, then the companion may receive a
kick. This kick may be imparted by either an asymmetric explosion, the Blaauw mechanism \citep{B61} or a combination of both. In this paper, we do not model the binary evolution and
kick velocities.  Rather we just assume that there are two populations,
one more numerous and does not receive kicks (the O stars), and one
that is less numerous and whose velocity distribution is dominated by
kicks.  

Once again, we may use our general analytic expression, relating the
separations, age and velocities, equation.~(\ref{eq:analytic2}), but this
time we express $v_L$ in terms of the age, $x = t_L/t_1$, and the
measured values of the separations,
\begin{equation}
\label{eq:vkickanalytic}
\frac{v_L}{v_O} = \frac{x_O}{x} \left [ 2 \left ( \frac{\langle S_{OL} \rangle }{\langle
  S_{O} \rangle }\right )^2
\frac{(1 - x^\tau)}{(1 -
  x_O^\tau)} - 1 \right ]^{1/2} \, .
\end{equation}

\citet{ST15} showed that the average distance from  LBVs to the nearest O
star is $\sim 6.5$ times larger than the average distance from O star
to the nearest O star. If the age of LBVs are similar to the average
mid-type O star, then in the assumption of the kick
model, this immediately implies that
$v_{\rm L}$ is roughly nine times larger than $v_{\rm O}$. In the next
subsection, we estimate the LBVs' drift velocity given this model.

\subsection{ESTIMATION AND INTERPRETATION OF THE KICK VELOCITY}
\label{kick velocity}
\begin{figure*}
\includegraphics[width=1.75\columnwidth]{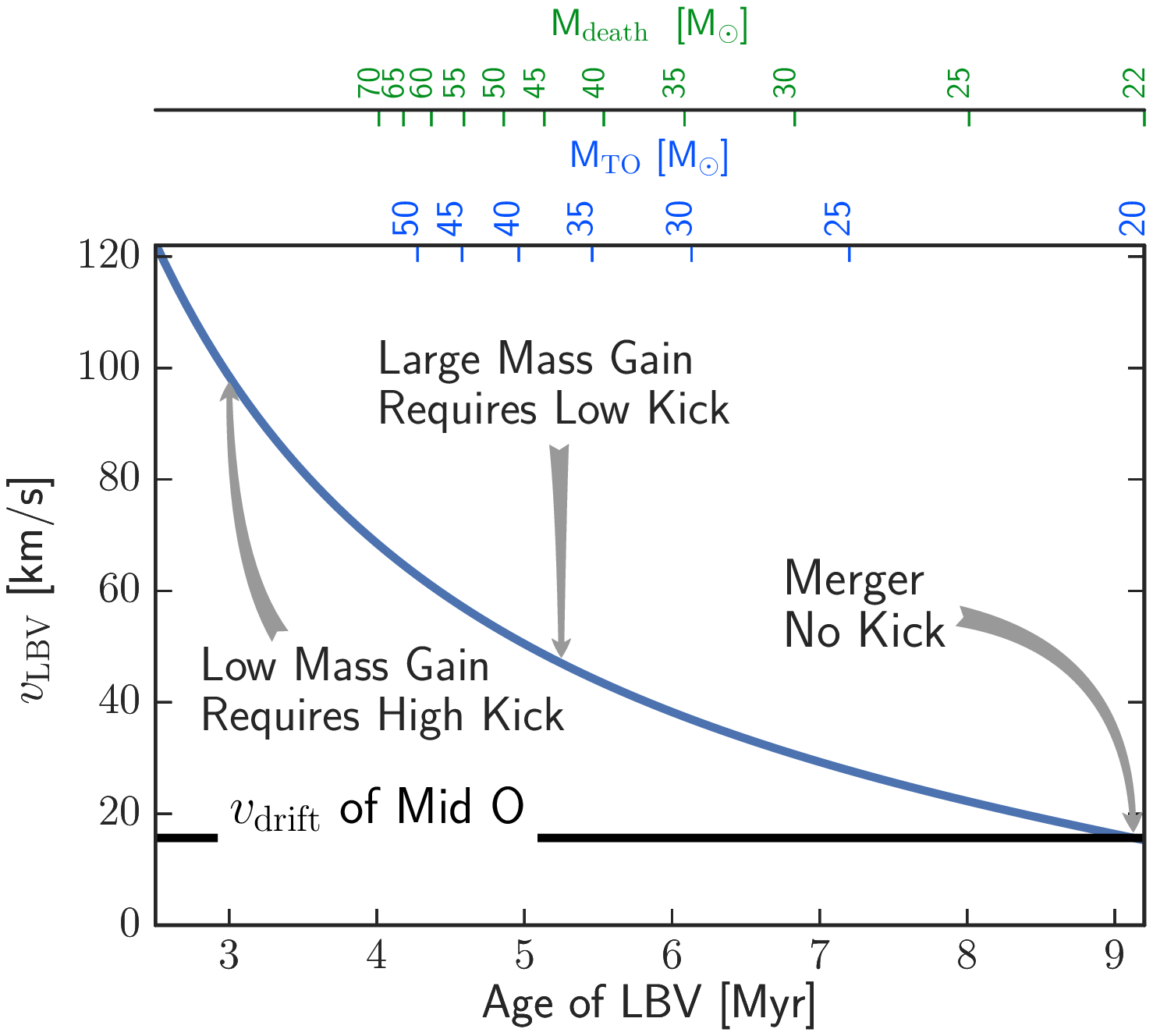}
\caption{LBV dispersion velocity as a function of LBV age.  Another binary model that is consistent with observations
  is one in which the LBV is a mass gainer and receives a kick when
  its primary companion explodes. The solid blue curve represents our
  analytic model, equation.~(\ref{eq:vkickanalytic}).  For reference, the solid black line shows the drift
  velocity of mid-type O stars, see Fig.~\ref{MeanVariance}.  As a
  mass gainer, the age of the LBV will be older than one would infer from
  its luminosity and mass.  If LBVs gain little to no mass, then the kick
  required to match the observed separations is in the range 0--105
  km/s. The lower end of this range corresponds to high mass transfer. Specifically, if LBVs have an age of the order of 9.2 Myr, then we
  suggest that LBVs are mergers and received no kick.}
\label{Kick}
\end{figure*}

Fig.~\ref{Kick} shows the inferred kick velocity as
a function of the LBV age.  If the mass gainer that eventually
becomes the LBV gains little mass,  then there is little discrepancy between the zero-age main-sequence mass and the final mass. In this case, there is little difference between its apparent age and its true main-sequence age. Then its true age is relatively short and the only way to get a large separation with a large kick velocity. In this scenario, we find that the kick can be as high as
105 km/s. If there is no mass gained and hence a larger kick (upper left in Fig.~\ref{Kick}) then the star that was kicked has not necessarily had any anomalous evolution (no accretion and spin-up) and hence gives no special explanation for its observed LBV instability. On the other hand, if the mass gain is high, then the true main-sequence age would be much older than the current mass implies. With a much older age, the velocity required to get a large separation is much lower.  It might even be zero, in which case, the LBV has gained so much mass that it is rejuvenated like a merger product. The horizontal black solid line represents the average
  observed separation for mid-type O stars. The solid blue line curve represents our model to infer the kick velocity, equation.~(\ref{eq:vkickanalytic}).

Though we predict that the kick velocities may be as high as $\sim$105
km/s, we note that the kick may be quite low, even near zero.
\citet{H16} argued that none of
the LBVs in the LMC have high
velocities. They suggest that most of the LBV velocities [listed in
table 3 of \citet{H16}] are consistent with the systemic
velocities of the LMC, concluding that the
observed velocities are inconsistent with the kick. In
Fig.~\ref{Kick}, we show that the kick velocity may be anywhere from
0 to $\sim$105 km/s depending on the orbital parameters at the time of the SN, and how much mass was transferred.  To further constrain the mass-gainer and
kick model, one will need to properly model binary evolution including
explosions and kicks in the context of dispersing cluster.  Current modelling efforts
 already indicate that the dispersal velocities from binary evolution could have a large range, even low dispersal velocities \citep{E11, M14,S16}. However, putting these binary models in the context of cluster dispersion is yet to be done. For
  now, we present the scale of the problem; in a subsequent paper, we
  will model the distribution of observed velocities one would
  expect.

\section{SUMMARY}\label{SUMMARY}
\citet{ST15} found that LBVs are surprisingly
isolated from other O stars. They suggested that the
relative isolation is inconsistent with a
single-star scenario in which the most massive stars
undergo an LBV phase on their way to evolving into a WR
star. Instead, they suggested that a
binary scenario is likely more consistent with the relative
  isolation of LBVs.  In this paper, we test this hypothesis by developing crude models for
  single-star and binary scenarios in the
  context of cluster dissolution. Even with
  these crude models, we find that the LBVs'
isolation is mostly inconsistent with
the standard passive single-star evolution model. In
  particular, if LBVs do evolve as single stars, then their isolation
  implies an age that is twice the maximum age of an average LBV. It may
  be the case that a small fraction of LBVs could evolve as single
  stars and still be consistent with the measured isolation. However,
  the fact that most LBVs are very isolated suggests that a large
  fraction is inconsistent with single-star evolution. For most LBVs,
there is a clear problem in the single-star model's mapping between
luminosity and kinematic age, and this is either because there is a
problem in the luminosity-to-mass mapping or there is a problem in
the mass-to-age mapping.  In this paper, we consider how binary
evolution might affect the latter, the mass-to-age mapping.
We find that the LBV
  isolation is most consistent with
 two binary scenarios: either LBVs are mass gainers
  and receive a kick anywhere from 0 to $\sim$105 km/s or they are
  the product of mergers and are rejuvenated stars. Of course, LBVs may actually represent a combination of these two scenarios. Based on their environments, it is quite possible that some are mass gainers and some are the product of mergers.
  
In order to constrain these models, we first
reproduce the results of \citet{ST15}.  Similarly, we obtain the
position of all O stars within $10^{\circ}$ projected radius of 30
Dor, and we construct
distributions of distances to the nearest O star. Our
distributions are very similar to theirs. We find that the distributions for the LBVs and O stars are very
unlikely to be drawn from the same parent
distributions. In particular, the average distance to the
nearest O star is $\sim$6.5 times larger for LBVs than O stars. To better inform
our models, we further characterize the
distributions  and find that all of the nearest neighbour distributions
are lognormal. The fact that the distributions are simple
  and lognormal demands that our models for cluster dissolution are also simple.

We propose simple Monte Carlo and analytic models for the
  dispersal of open clusters of O stars. In this model, we sample from distributions of cluster
sizes, the Salpeter IMF for stars and
random ages. To match the observed separations for early O
  stars, we find that the early-type
  clusters need a drift velocity of the order of 7
km/s.  For the mid and late types, we
require drift velocities of the order
of 15 and 11 km/s, respectively. The higher
  drift velocities for later type O stars hint that binarity and
  kicks may play a prominent role in cluster dissolution. In fact, some fraction of later type O stars may be mass gainers or the product of mergers. In a future paper,
  we will investigate whether one can constrain the fraction of kicks
  and strong binary interaction.

Using the
  results of the Monte Carlo simulations as a guide, we develop an
analytical model for the average separation
as a function of drift velocity and time. These analytic scalings
  strongly suggest that
LBV isolation is inconsistent with single-star stellar
evolution. Instead, these scalings in combination with LBV isolation suggest that either
LBVs have lower initial masses (and hence longer lifetimes) than one would infer from luminosities or the isolation is most consistent with some sort of binary interaction:
either a merger or kick. If LBVs have the same
dispersion velocity that we infer from mid-type O stars,
then the time to get to the relatively large isolation is 9.2
Myr. However, the
average mass of LBVs is 50 $\msun$ which has a maximum
time of $\sim$4.8 Myr. This is clearly
inconsistent. On the other hand, binary
interactions can easily achieve large isolations. In one scenario,
LBVs might be the product of
the merger of two massive O-type stars, in which the primary has a
mass of at least about 19 $\msun$.  Another possibility
is a kick due to binary evolution. In this binary
scenario, the less massive stars (pre-LBV stars) gain mass
from its companion. After mass transfer, the primary explodes
as an SN and the LBV
receives a kick anywhere from 0 to $\sim$105 km/s.

With current observations and theory, either binary model is
consistent with the data.  To further constrain which binary model is
most consistent, we need to gather more data and develop better models. For example, detailed
  kinematic observations and theory would help to distinguish between
  these two models. \citet{H16} suggest that the velocities of LBVs are too
  low to be consistent with the kick scenario. However, there are binary scenarios that would produce low kick velocities. For example, if the secondary accretes so much mass that it becomes the more massive star in the binary, then this much more massive secondary will have a low orbital velocity in its binary orbit. When the low-mass primary explodes, the mass gainer drifts away at its low orbital speed. Hence, \citet{S16} pointed out that large kick speeds are not necessarily expected, especially when there has been a significant amount of mass gained \citep{E11, M14}. We show that
  the mass-gainer scenario currently predicts a wide range of kick
  velocities. To truly test the consistency of the kick model, we must first
model binary evolution and develop a model for the appearance of the kinematics, including
randomness, projection, etc. The merger model would
  manifest as an inconsistency between the maximum age of the LBV and
  the surrounding stellar population.  Therefore, to constrain the merger
  model, we need better mass estimates for the LBVs and age estimates
  for the surrounding stellar populations.
  
In conclusion, we develop models for cluster dissolution and the spatial distribution
  of LBVs and O stars. These models
suggest that single-star evolution in passively
evolving clusters is inconsistent with the extreme isolation
  of LBVs. Instead, we find that either LBVs are less massive than
  their luminosities would imply or binary
interaction is most consistent with LBV
  isolation. In particular, we crudely find that two binary
  scenarios are consistent with the data.  Either LBVs are mass
  gainers and received a kick when the primary exploded or they are
  rejuvenated stars, being the product of mergers.

\section*{ACKNOWLEDGMENTS}
This research has made use of the SIMBAD data base, operated at CDS, Strasbourg, France. Support for NS was provided by the National Science Foundation (NSF) through grants AST-1312221 and AST-1515559 to the University of Arizona.

 \bibliography{Mojgan}

\begin{thebibliography}{}
\makeatletter
\relax
\def\mn@urlcharsother{\let\do\@makeother \do\$\do\&\do\#\do\^\do\_\do\%\do\~}
\def\mn@doi{\begingroup\mn@urlcharsother \@ifnextchar [ {\mn@doi@}
  {\mn@doi@[]}}
\def\mn@doi@[#1]#2{\def\@tempa{#1}\ifx\@tempa\@empty \href
  {http://dx.doi.org/#2} {doi:#2}\else \href {http://dx.doi.org/#2} {#1}\fi
  \endgroup}
\def\mn@eprint#1#2{\mn@eprint@#1:#2::\@nil}
\def\mn@eprint@arXiv#1{\href {http://arxiv.org/abs/#1} {{\tt arXiv:#1}}}
\def\mn@eprint@dblp#1{\href {http://dblp.uni-trier.de/rec/bibtex/#1.xml}
  {dblp:#1}}
\def\mn@eprint@#1:#2:#3:#4\@nil{\def\@tempa {#1}\def\@tempb {#2}\def\@tempc
  {#3}\ifx \@tempc \@empty \let \@tempc \@tempb \let \@tempb \@tempa \fi \ifx
  \@tempb \@empty \def\@tempb {arXiv}\fi \@ifundefined
  {mn@eprint@\@tempb}{\@tempb:\@tempc}{\expandafter \expandafter \csname
  mn@eprint@\@tempb\endcsname \expandafter{\@tempc}}}

\bibitem[\protect\citeauthoryear{{Blaauw}}{{Blaauw}}{1961}]{B61}
{Blaauw} A.,  1961, \bain, \href
  {http://adsabs.harvard.edu/abs/1961BAN....15..265B} {15, 265}

\bibitem[\protect\citeauthoryear{{Bohannan} \& {Walborn}}{{Bohannan} \&
  {Walborn}}{1989}]{BW89}
{Bohannan} B.,  {Walborn} N.~R.,  1989, \mn@doi [\pasp] {10.1086/132463}, \href
  {http://adsabs.harvard.edu/abs/1989PASP..101..520B} {101, 520}

\bibitem[\protect\citeauthoryear{{Eldridge}, {Langer}  \& {Tout}}{{Eldridge}
  et~al.}{2011}]{E11}
{Eldridge} J.~J.,  {Langer} N.,   {Tout} C.~A.,  2011, \mn@doi [\mnras]
  {10.1111/j.1365-2966.2011.18650.x}, \href
  {http://adsabs.harvard.edu/abs/2011MNRAS.414.3501E} {414, 3501}

\bibitem[\protect\citeauthoryear{{Elmegreen} \& {Efremov}}{{Elmegreen} \&
  {Efremov}}{1997}]{E97}
{Elmegreen} B.~G.,  {Efremov} Y.~N.,  1997, \apj, \href
  {http://adsabs.harvard.edu/abs/1997ApJ...480..235E} {480, 235}

\bibitem[\protect\citeauthoryear{{Gallagher}}{{Gallagher}}{1989}]{G89}
{Gallagher} J.~S.,  1989, in {Davidson} K.,  {Moffat} A.~F.~J.,   {Lamers}
  H.~J.~G.~L.~M.,  eds,  Astrophysics and Space Science Library Vol. 157, IAU
  Colloq. 113: Physics of Luminous Blue Variables. pp 185--192,
  \mn@doi{10.1007/978-94-009-1031-7_22}

\bibitem[\protect\citeauthoryear{{Gr{\"a}fener}, {Owocki}  \&
  {Vink}}{{Gr{\"a}fener} et~al.}{2012}]{G12}
{Gr{\"a}fener} G.,  {Owocki} S.~P.,   {Vink} J.~S.,  2012, \mn@doi [\aap]
  {10.1051/0004-6361/201117497}, \href
  {http://adsabs.harvard.edu/abs/2012A%26A...538A..40G} {538, A40}

\bibitem[\protect\citeauthoryear{{Groh}, {Hillier}, {Damineli}, {Whitelock},
  {Marang}  \& {Rossi}}{{Groh} et~al.}{2009}]{G09}
{Groh} J.~H.,  {Hillier} D.~J.,  {Damineli} A.,  {Whitelock} P.~A.,  {Marang}
  F.,   {Rossi} C.,  2009, \mn@doi [\apj] {10.1088/0004-637X/698/2/1698}, \href
  {http://adsabs.harvard.edu/abs/2009ApJ...698.1698G} {698, 1698}

\bibitem[\protect\citeauthoryear{{Humphreys} \& {Davidson}}{{Humphreys} \&
  {Davidson}}{1994}]{HD94}
{Humphreys} R.~M.,  {Davidson} K.,  1994, \mn@doi [\pasp] {10.1086/133478},
  \href {http://adsabs.harvard.edu/abs/1994PASP..106.1025H} {106, 1025}

\bibitem[\protect\citeauthoryear{{Humphreys}, {Weis}, {Davidson}  \&
  {Gordon}}{{Humphreys} et~al.}{2016}]{H16}
{Humphreys} R.~M.,  {Weis} K.,  {Davidson} K.,   {Gordon} M.~S.,  2016,
  preprint, \href {http://adsabs.harvard.edu/abs/2016arXiv160301278H} {}
  (\mn@eprint {arXiv} {1603.01278})

\bibitem[\protect\citeauthoryear{{Ivezic}, {J. Connolly}, {T VanderPlas}  \&
  {Gra}}{{Ivezic} et~al.}{2014}]{Z14}
{Ivezic} Z.,  {J. Connolly} A.~.,  {T VanderPlas} J.~.,   {Gra} A.,  2014,
  Statistics, Data Mining, and Machine Learning in Astronomy.
Princeton Series in Modern Observational Astronomy, Princeton University Press

\bibitem[\protect\citeauthoryear{{Izzard}, {Tout}, {Karakas}  \&
  {Pols}}{{Izzard} et~al.}{2004}]{I04}
{Izzard} R.~G.,  {Tout} C.~A.,  {Karakas} A.~I.,   {Pols} O.~R.,  2004, \mn@doi
  [\mnras] {10.1111/j.1365-2966.2004.07446.x}, \href
  {http://cdsads.u-strasbg.fr/abs/2004MNRAS.350..407I} {350, 407}

\bibitem[\protect\citeauthoryear{{Izzard}, {Dray}, {Karakas}, {Lugaro}  \&
  {Tout}}{{Izzard} et~al.}{2006}]{I06}
{Izzard} R.~G.,  {Dray} L.~M.,  {Karakas} A.~I.,  {Lugaro} M.,   {Tout} C.~A.,
  2006, \mn@doi [\aap] {10.1051/0004-6361:20066129}, \href
  {http://cdsads.u-strasbg.fr/abs/2006A%26A...460..565I} {460, 565}

\bibitem[\protect\citeauthoryear{{Izzard}, {Glebbeek}, {Stancliffe}  \&
  {Pols}}{{Izzard} et~al.}{2009}]{I09}
{Izzard} R.~G.,  {Glebbeek} E.,  {Stancliffe} R.~J.,   {Pols} O.~R.,  2009,
  \mn@doi [\aap] {10.1051/0004-6361/200912827}, \href
  {http://cdsads.u-strasbg.fr/abs/2009A%26A...508.1359I} {508, 1359}

\bibitem[\protect\citeauthoryear{{Justham}, {Podsiadlowski}  \&
  {Vink}}{{Justham} et~al.}{2014}]{J14}
{Justham} S.,  {Podsiadlowski} P.,   {Vink} J.~S.,  2014, \mn@doi [\apj]
  {10.1088/0004-637X/796/2/121}, \href
  {http://adsabs.harvard.edu/abs/2014ApJ...796..121J} {796, 121}

\bibitem[\protect\citeauthoryear{{Klein}, {Cenko}, {Miller}, {Norman}  \&
  {Bloom}}{{Klein} et~al.}{2014}]{K14}
{Klein} C.~R.,  {Cenko} S.~B.,  {Miller} A.~A.,  {Norman} D.~J.,   {Bloom}
  J.~S.,  2014, preprint, \href
  {http://adsabs.harvard.edu/abs/2014arXiv1405.1035K} {} (\mn@eprint {arXiv}
  {1405.1035})

\bibitem[\protect\citeauthoryear{{Kotak} \& {Vink}}{{Kotak} \&
  {Vink}}{2006}]{KV06}
{Kotak} R.,  {Vink} J.~S.,  2006, \mn@doi [\aap] {10.1051/0004-6361:20065800},
  \href {http://adsabs.harvard.edu/abs/2006A%26A...460L...5K} {460, L5}

\bibitem[\protect\citeauthoryear{{Krumholz} \& {Tan}}{{Krumholz} \&
  {Tan}}{2007}]{K07}
{Krumholz} M.~R.,  {Tan} J.~C.,  2007, \mn@doi [\apj] {10.1086/509101}, \href
  {http://adsabs.harvard.edu/abs/2007ApJ...654..304K} {654, 304}

\bibitem[\protect\citeauthoryear{{Kudritzki} \& {Puls}}{{Kudritzki} \&
  {Puls}}{2000}]{K00}
{Kudritzki} R.-P.,  {Puls} J.,  2000, \mn@doi [\araa]
  {10.1146/annurev.astro.38.1.613}, \href
  {http://adsabs.harvard.edu/abs/2000ARA%26A..38..613K} {38, 613}

\bibitem[\protect\citeauthoryear{{Ma{\'{\i}}z Apell{\'a}niz}
  et~al.,}{{Ma{\'{\i}}z Apell{\'a}niz} et~al.}{2013}]{M13}
{Ma{\'{\i}}z Apell{\'a}niz} J.,  et~al., 2013, in Massive Stars: From alpha to
  Omega. p.~198, \url {http://a2omega-conference.net}

\bibitem[\protect\citeauthoryear{{Marconi} \& {Clementini}}{{Marconi} \&
  {Clementini}}{2005}]{MC05}
{Marconi} M.,  {Clementini} G.,  2005, \mn@doi [\aj] {10.1086/429525}, \href
  {http://adsabs.harvard.edu/abs/2005AJ....129.2257M} {129, 2257}

\bibitem[\protect\citeauthoryear{{Martins} \& {Palacios}}{{Martins} \&
  {Palacios}}{2013}]{MP13}
{Martins} F.,  {Palacios} A.,  2013, \mn@doi [\aap]
  {10.1051/0004-6361/201322480}, \href
  {http://adsabs.harvard.edu/abs/2013A%26A...560A..16M} {560, A16}

\bibitem[\protect\citeauthoryear{{Martins}, {Schaerer}  \& {Hillier}}{{Martins}
  et~al.}{2005}]{M05}
{Martins} F.,  {Schaerer} D.,   {Hillier} D.~J.,  2005, \mn@doi [\aap]
  {10.1051/0004-6361:20042386}, \href
  {http://adsabs.harvard.edu/abs/2005A%26A...436.1049M} {436, 1049}

\bibitem[\protect\citeauthoryear{{Meynet} \& {Maeder}}{{Meynet} \&
  {Maeder}}{2005}]{MM05}
{Meynet} G.,  {Maeder} A.,  2005, \mn@doi [\aap] {10.1051/0004-6361:20047106},
  \href {http://adsabs.harvard.edu/abs/2005A%26A...429..581M} {429, 581}

\bibitem[\protect\citeauthoryear{{Ofek} et~al.,}{{Ofek} et~al.}{2013}]{O13}
{Ofek} E.~O.,  et~al., 2013, \mn@doi [\nat] {10.1038/nature11877}, \href
  {http://adsabs.harvard.edu/abs/2013Natur.494...65O} {494, 65}

\bibitem[\protect\citeauthoryear{{Owocki}, {Gayley}  \& {Shaviv}}{{Owocki}
  et~al.}{2004}]{O04}
{Owocki} S.~P.,  {Gayley} K.~G.,   {Shaviv} N.~J.,  2004, \mn@doi [\apj]
  {10.1086/424910}, \href {http://adsabs.harvard.edu/abs/2004ApJ...616..525O}
  {616, 525}

\bibitem[\protect\citeauthoryear{{Puls}, {Vink}  \& {Najarro}}{{Puls}
  et~al.}{2008}]{P08}
{Puls} J.,  {Vink} J.~S.,   {Najarro} F.,  2008, \mn@doi [\aapr]
  {10.1007/s00159-008-0015-8}, \href
  {http://adsabs.harvard.edu/abs/2008A%26ARv..16..209P} {16, 209}

\bibitem[\protect\citeauthoryear{{Smith}}{{Smith}}{2014}]{S14}
{Smith} N.,  2014, \mn@doi [\araa] {10.1146/annurev-astro-081913-040025}, \href
  {http://adsabs.harvard.edu/abs/2014ARA%26A..52..487S} {52, 487}

\bibitem[\protect\citeauthoryear{{Smith}}{{Smith}}{2016}]{S16}
{Smith} N.,  2016, \mn@doi [\mnras] {10.1093/mnras/stw1533}, \href
  {http://adsabs.harvard.edu/abs/2016MNRAS.461.3353S} {461, 3353}

\bibitem[\protect\citeauthoryear{{Smith} \& {Owocki}}{{Smith} \&
  {Owocki}}{2006}]{SO06}
{Smith} N.,  {Owocki} S.~P.,  2006, \mn@doi [\apjl] {10.1086/506523}, \href
  {http://adsabs.harvard.edu/abs/2006ApJ...645L..45S} {645, L45}

\bibitem[\protect\citeauthoryear{{Smith} \& {Stassun}}{{Smith} \&
  {Stassun}}{2017}]{SS17}
{Smith} N.,  {Stassun} K.~G.,  2017, \mn@doi [\aj] {10.3847/1538-3881/aa5d0c},
  \href {http://adsabs.harvard.edu/abs/2017AJ....153..125S} {153, 125}

\bibitem[\protect\citeauthoryear{{Smith} \& {Tombleson}}{{Smith} \&
  {Tombleson}}{2015}]{ST15}
{Smith} N.,  {Tombleson} R.,  2015, \mn@doi [\mnras] {10.1093/mnras/stu2430},
  \href {http://adsabs.harvard.edu/abs/2015MNRAS.447..598S} {447, 598}

\bibitem[\protect\citeauthoryear{{Smith} et~al.,}{{Smith} et~al.}{2007}]{S07}
{Smith} N.,  et~al., 2007, \mn@doi [\apj] {10.1086/519949}, \href
  {http://adsabs.harvard.edu/abs/2007ApJ...666.1116S} {666, 1116}

\bibitem[\protect\citeauthoryear{{Smith} et~al.,}{{Smith} et~al.}{2008}]{S08}
{Smith} N.,  et~al., 2008, \mn@doi [\apj] {10.1086/590141}, \href
  {http://adsabs.harvard.edu/abs/2008ApJ...686..485S} {686, 485}

\bibitem[\protect\citeauthoryear{{Smith}, {Li}, {Silverman}, {Ganeshalingam}
  \& {Filippenko}}{{Smith} et~al.}{2011}]{S11}
{Smith} N.,  {Li} W.,  {Silverman} J.~M.,  {Ganeshalingam} M.,   {Filippenko}
  A.~V.,  2011, \mn@doi [\mnras] {10.1111/j.1365-2966.2011.18763.x}, \href
  {http://adsabs.harvard.edu/abs/2011MNRAS.415..773S} {415, 773}

\bibitem[\protect\citeauthoryear{{Vink}}{{Vink}}{2012}]{V12}
{Vink} J.~S.,  2012, in {Davidson} K.,  {Humphreys} R.~M.,  eds,  Astrophysics
  and Space Science Library Vol. 384, Eta Carinae and the Supernova Impostors.
  p.~221 (\mn@eprint {arXiv} {0905.3338}),
  \mn@doi{10.1007/978-1-4614-2275-4_10}

\bibitem[\protect\citeauthoryear{{Vink}, {de Koter}  \& {Lamers}}{{Vink}
  et~al.}{2001}]{VD01}
{Vink} J.~S.,  {de Koter} A.,   {Lamers} H.~J.~G.~L.~M.,  2001, \mn@doi [\aap]
  {10.1051/0004-6361:20010127}, \href
  {http://adsabs.harvard.edu/abs/2001A%26A...369..574V} {369, 574}

\bibitem[\protect\citeauthoryear{{Walborn}}{{Walborn}}{1977}]{W77}
{Walborn} N.~R.,  1977, \mn@doi [\apj] {10.1086/155334}, \href
  {http://adsabs.harvard.edu/abs/1977ApJ...215...53W} {215, 53}

\bibitem[\protect\citeauthoryear{{Walker}}{{Walker}}{2012}]{W12}
{Walker} A.~R.,  2012, \mn@doi [\apss] {10.1007/s10509-011-0961-x}, \href
  {http://adsabs.harvard.edu/abs/2012Ap%26SS.341...43W} {341, 43}

\bibitem[\protect\citeauthoryear{{Weis}}{{Weis}}{2003}]{W03}
{Weis} K.,  2003, \mn@doi [\aap] {10.1051/0004-6361:20030921}, \href
  {http://adsabs.harvard.edu/abs/2003A%26A...408..205W} {408, 205}

\bibitem[\protect\citeauthoryear{{Wolf}}{{Wolf}}{1989}]{W89}
{Wolf} B.,  1989, \aap, \href
  {http://adsabs.harvard.edu/abs/1989A%26A...217...87W} {217, 87}

\bibitem[\protect\citeauthoryear{{Woosley} \& {Heger}}{{Woosley} \&
  {Heger}}{2015}]{W15}
{Woosley} S.~E.,  {Heger} A.,  2015, in {Vink} J.~S.,  ed.,  Astrophysics and
  Space Science Library Vol. 412, Very Massive Stars in the Local Universe.
  p.~199 (\mn@eprint {arXiv} {1406.5657}), \mn@doi{10.1007/978-3-319-09596-7_7}

\bibitem[\protect\citeauthoryear{{Woosley}, {Heger}  \& {Weaver}}{{Woosley}
  et~al.}{2002}]{W02}
{Woosley} S.~E.,  {Heger} A.,   {Weaver} T.~A.,  2002, \mn@doi [Reviews of
  Modern Physics] {10.1103/RevModPhys.74.1015}, \href
  {http://adsabs.harvard.edu/abs/2002RvMP...74.1015W} {74, 1015}

\bibitem[\protect\citeauthoryear{{de Koter}, {Lamers}  \& {Schmutz}}{{de Koter}
  et~al.}{1996}]{D96}
{de Koter} A.,  {Lamers} H.~J.~G.~L.~M.,   {Schmutz} W.,  1996, \aap, \href
  {http://adsabs.harvard.edu/abs/1996A%26A...306..501D} {306, 501}

\bibitem[\protect\citeauthoryear{{de Mink}, {Sana}, {Langer}, {Izzard}  \&
  {Schneider}}{{de Mink} et~al.}{2014}]{M14}
{de Mink} S.~E.,  {Sana} H.,  {Langer} N.,  {Izzard} R.~G.,   {Schneider}
  F.~R.~N.,  2014, \mn@doi [\apj] {10.1088/0004-637X/782/1/7}, \href
  {http://adsabs.harvard.edu/abs/2014ApJ...782....7D} {782, 7}

\bibitem[\protect\citeauthoryear{{van Genderen}}{{van Genderen}}{2001}]{V01}
{van Genderen} A.~M.,  2001, \mn@doi [\aap] {10.1051/0004-6361:20000022}, \href
  {http://adsabs.harvard.edu/abs/2001A%26A...366..508V} {366, 508}

\makeatother
\end{thebibliography}

\label{lastpage}
 \end{document}